{}%disable hyper at arxiv
{}%use pdflatex at arxiv

\def\verdraft{3}
\def\verpublish{4}
\def\verarxiv{5}

\def\version{5}

\documentclass[12pt,a4paper]{article}

\usepackage{amsmath,amssymb}
\ifnum\version<2\usepackage{showkeys}\fi
\usepackage{graphicx}
\usepackage[bookmarks=true,hyperfigures=true]{hyperref}
\usepackage[usenames,dvipsnames]{color}
\usepackage[nosort]{cite}
\usepackage[bulletsep]{collref}
\usepackage[utf8]{inputenc}

\def\showkeysrefformat#1{{\normalfont\tiny\ttfamily#1}}
\makeatletter
\def\SK@@ref#1>#2\SK@{%
 {\@inlabelfalse\leavevmode\vbox to\z@{%
 \vss\SK@refcolor\rlap{\vrule\raise .75em%
  \hbox{\showkeysrefformat{#2}}}}}}
\makeatother

\usepackage[a4paper,text={160mm,247mm},centering]{geometry}
\RequirePackage{verbatim}

\makeatletter
\ifnum\version<\verarxiv\relax
\AddToHook{enddocument/afteraux}{\immediate\write18{bibtex \jobname}}
\newwrite\bibinl@out
\newenvironment{bibtex}[1][\jobname]{%
  \immediate\openout\bibinl@out #1.bib
  \immediate\write\bibinl@out{\@percentchar generated from `\jobname' starting line \the\inputlineno^^J}%
  \def\verbatim@processline{\immediate\write\bibinl@out{\the\verbatim@line}}%
  \@bsphack\let\do\@makeother\dospecials\catcode`\^^M\active\verbatim@start
}%
{\immediate\closeout\bibinl@out\@esphack}
\else
\newenvironment{bibtex}[1][\jobname]{\comment}{\endcomment}
\fi
\makeatother

\allowdisplaybreaks[3]

\numberwithin{equation}{section}

\usepackage[font=small,labelfont=bf,width=0.85\textwidth]{caption}

\newcommand{\alignrel}[1][=]{\mathrel{\phantom{#1}}{}}
\makeatletter
\newcommand{\eqsep}{\@ifstar{\hspace{1em}}{\hspace{2em minus 1em}}}
\newcommand{\@eqjoin}[2]{\hspace{#1}#2\hspace{#1}}
\newcommand{\eqjoin}{\@ifstar{\@eqjoin{1em}}{\@eqjoin{2em minus 1em}}}
\makeatother

\usepackage{mathtools}

\newcommand{\mycomment}[1]{}

\renewcommand{\minalignsep}{2em}

\makeatletter
\expandafter\def\expandafter\calc@shift@align\expandafter
  {\expandafter\hook@align@sep\calc@shift@align}
\def\hook@align@sep{\ifnum\xatlevel@=\@ne
  \dimen@\displaywidth
  \advance\dimen@-\totwidth@
  \@tempcntb\maxfields@
  \divide\@tempcntb\tw@
  \@tempcnta\@tempcntb
  \advance\@tempcntb\m@ne
  \global\eqnshift@\dimen@
  \global\alignsep@\minalignsep\relax
  \global\advance\eqnshift@-\@tempcntb\alignsep@
  \global\divide\eqnshift@\tw@
  \ifdim\eqnshift@<\z@
   \global\eqnshift@\z@
  \fi
 \fi}
\makeatother

\makeatletter
\expandafter\def\expandafter\align\expandafter
  {\expandafter\hook@align@\align}
\expandafter\def\expandafter\math@cr@@@align\expandafter
  {\expandafter\hook@align@cr\math@cr@@@align}
\expandafter\def\expandafter\endalign\expandafter
  {\expandafter\hook@align@end\endalign}
\expandafter\def\expandafter\measure@\expandafter#\expandafter1\expandafter
  {\measure@{#1}\hook@align@}
\def\hook@align@{\gdef\hook@align@end{\donumber}%
  \gdef\hook@align@cr{\nonumber}}
\def\hook@align@cr{}
\def\hook@align@end{}

\newcommand{\donumber}{\gdef\hook@align@cr{\gdef\hook@align@cr{\nonumber}}}%
\newcommand{\numberhere}{\donumber\gdef\hook@align@end{}}%
\makeatother

\usepackage{mathfixs}
\ProvideMathFix{autobold,greekcaps,frac,root}
\ProvideMathFix{multskip}

\RequirePackage[extdef]{delimset}

\ProvideMathFix{vfrac,rfrac}
\newcommand{\iunit}{\mathring{\imath}}

\newcommand{\half}{\rfrac{1}{2}}

\newcommand{\alg}[1]{\mathfrak{#1}}

\newcommand{\gen}[1][J]{\mathrm{#1}{}}
\newcommand{\yanghat}[1]{\widehat{#1}}
\newcommand{\genyang}[1][J]{\yanghat{\gen[#1]}{}}
\newcommand{\gencart}{\gen[T]}
\newcommand{\chargecart}{t}
\newcommand{\swdl}[1]{{(#1)}}

\newcommand{\shift}{\gen[U]}
\newcommand{\invact}{\oper{X}}
\newcommand{\invactyang}{\yanghat{\invact}}
\newcommand{\inveom}{\oper{Y}}
\newcommand{\inveomyang}{\yanghat{\inveom}}
\newcommand{\twistmap}{\gen[F]}
\newcommand{\twist}{\twistmap}
\newcommand{\twiststar}{\star}
\newcommand{\gentwist}{\gen[K]}
\newcommand{\gentwistyang}{\yanghat{\gentwist}}

\newcommand{\superN}{\mathcal{N}}

\newcommand{\oper}[1]{\mathcal{#1}}

\newcommand{\action}{\oper{S}}
\newcommand{\field}{Z}
\newcommand{\eomfor}[1]{\breve{#1}}

\newcommand{\var}{\delta}

\newcommand{\nfield}[1]{{[#1]}}

\newcommand{\der}{\mathrm{d}}
\newcommand{\diff}[2][.]{\mathinner{\der#2\if #1.\else^{#1}\fi}}
\newcommand{\defeq}{\coloneqq}

\IfFileExists{dsfont.sty}{\RequirePackage{dsfont}}{\RequirePackage{bbm}}

\DeclareMathOperator{\tr}{tr}

\ifdefined\mathds\providecommand{\idop}{\mathds{1}}\fi
\ifdefined\mathbbm\providecommand{\idop}{\mathbbm{1}}\fi
\providecommand{\idop}{1}

\def\[{\begin{equation}}
\def\]{\end{equation}}

\providecommand{\href}[2]{#2}

\makeatletter
\def\mr@ignsp#1 {\ifx\:#1\@empty\else #1\expandafter\mr@ignsp\fi}%
\newcommand{\multiref}[1]{\begingroup%\let\protect\string%
\xdef\mr@no@sparg{\expandafter\mr@ignsp#1 \: }%
\def\mr@comma{}%
\@for\mr@refs:=\mr@no@sparg\do{\mr@comma\def\mr@comma{,}\ref{\mr@refs}}%
\endgroup}
\renewcommand{\eqref}[1]{(\multiref{#1})}
\makeatother

\makeatletter
\newcommand{\namedref}[2]{\hyperref[#2]{#1~\ref*{#2}}}
\newcommand{\secref}{\@ifstar{\namedref{Section}}{\namedref{Sec.}}}
\newcommand{\appref}{\@ifstar{\namedref{Appendix}}{\namedref{App.}}}
\newcommand{\tabref}{\@ifstar{\namedref{Table}}{\namedref{Tab.}}}
\newcommand{\figref}{\@ifstar{\namedref{Figure}}{\namedref{Fig.}}}
\makeatother

\let\oldbib=\thebibliography
\def\thebibliography{\phantomsection\addcontentsline{toc}{section}{\refname}\oldbib}

\let\oldtoc=\tableofcontents
\def\tableofcontents{\phantomsection\addcontentsline{toc}{section}{\contentsname}\oldtoc}

\providecommand{\hypersetup}[1]{}

\hypersetup{plainpages=false}
\hypersetup{pdfpagemode=UseNone}
\hypersetup{bookmarksnumbered=true}
\hypersetup{pdfstartview=FitH}
\hypersetup{colorlinks=false}
\hypersetup{citebordercolor={.5 1 .5}}
\hypersetup{urlbordercolor={.5 1 1}}
\hypersetup{linkbordercolor={1 .7 .7}}
\ifnum\version<\verpublish\PassOptionsToPackage{draft}{metastr}\fi
\RequirePackage{metastr}

\ifnum\version<\verdraft\metasetterm{draft}{EDITING}\fi

\usepackage{ifpdf}
\ifpdf\else\RequirePackage[active]{srcltx}\fi
\RequirePackage{color}

\newcommand{\remark}[2][]{}

\ifnum\version<\verpublish
\renewcommand{\remark}[2][]{{%
  \def\emph{\textsl}\def\textbullet{$\bullet$}%
  \def\tmparga{#1}\def\tmpargb{draft}\ifx\tmparga\tmpargb%
  \color[rgb]{0.5,0,0}\normalfont\sffamily\hspace{1ex} #2\hspace{1ex}\fi}}
\fi

\ifnum\version<\verdraft
\renewcommand{\remark}[2][]{{\normalfont\sffamily\hspace{1ex}%
  \def\emph{\textsl}\def\textbullet{$\bullet$}%
  \def\tmparga{#1}%
  \def\tmpargb{NB}\ifx\tmparga\tmpargb\color[rgb]{0,0,0.8}\fi%
  \def\tmpargb{BK}\ifx\tmparga\tmpargb\color[rgb]{0,0.5,0}\fi%
  \def\tmpargb{}\ifx\tmparga\tmpargb\color{red}\fi%
  \def\tmpargb{}\ifx\tmparga\tmpargb\else \textbf{#1:} \fi%
  #2\hspace{1ex}}}
\fi

\metaset{title}{Yangian Form-alism for Planar Gauge Theories}
\metaset{author}{Niklas Beisert, Benedikt König}
\metaset[print]{author}{Niklas Beisert\textsuperscript{1}, Benedikt König\textsuperscript{2}}

\begin{document}

\ifnum\version>0
\pdfbookmark[1]{Title Page}{title}
\thispagestyle{empty}

\begingroup\raggedleft\footnotesize\ttfamily
%\arxivlink{yymm.nnnnn}
%AEI-....
\par\endgroup

\vspace*{2cm}
\begin{center}%
\begingroup\Large\bfseries\metapick[print]{title}\par\endgroup
\vspace{1cm}

\begingroup\scshape
\metapick[print]{author}
\endgroup
\vspace{5mm}

\textsuperscript{1} 
\textit{Institut für Theoretische Physik,\\
Eidgenössische Technische Hochschule Zürich,\\
Wolfgang-Pauli-Strasse 27, 8093 Zürich, Switzerland}
\par\vspace{1mm}
\texttt{nbeisert@itp.phys.ethz.ch}
\vspace{5mm}

\textsuperscript{2}
\textit{Max-Planck-Institut für Gravitationsphysik (AEI),\\
Am Mühlenberg 1, DE-14476 Potsdam, Germany}
\par\vspace{1mm}
\texttt{benedikt.koenig@aei.mpg.de}
\vspace{8mm}

\vfill

\metaif[]{draft}{
\vspace{5mm}
\begingroup\bfseries\Large \metaterm{draft}\endgroup
\par
\vspace{5mm}
\vfill
}{}

\textbf{Abstract}\vspace{5mm}

\begin{minipage}{12.7cm}
In this article, we reconsider the formulation of Yangian symmetry for planar $\superN=4$ supersymmetric Yang--Mills theory,
and we investigate to what extent this symmetry lifts to the beta/gamma-deformation of the model.
We first apply cohomology of variational forms
towards a thorough derivation of the invariance statement for the undeformed action
from covariance of the equations of motion under the Yangian algebra.
We then apply a twist deformation to these statements
paying particular attention to cyclicity aspects.
We find that the equations of motion remain covariant
while invariance of the action only holds 
for the Yangian subalgebra that is uncharged under the twist.
\end{minipage}

\vspace*{3cm}

\end{center}

\ifnum\version>1
\newpage
\tableofcontents
\fi

\newpage
\fi

%%%%%%%%%%%%%%%%%%%%%%%%%%%%%%%%%%%%%%%%%%%%%%%%%%%%%%%%%%%%%%%%%%%%%%%%%%%%%%%%
%%%%%%%%%%%%%%%%%%%%%%%%%%%%%%%%%%%%%%%%%%%%%%%%%%%%%%%%%%%%%%%%%%%%%%%%%%%%%%%%
\section{Introduction}
\label{sec:intro}

Integrability is a powerful feature of certain theoretical physics models
allowing to evaluate relevant observables conveniently and exactly.
For $\superN=4$ supersymmetric Yang--Mills theory ($\superN=4$ SYM)
and $\superN=6$ supersymmetric Chern--Simons theory (also known as ABJ(M)) in the planar limit,
integrability methods have proved of extreme value
towards the exact evaluation of many observables,
which in itself is strong evidence for integrability of  these models \cite{Beisert:2010jr}. 
However, the established notions of integrability do not fully apply here,
and a clear-cut definition for integrability of these models,
by which the above powerful methods could be justified and derived,
is still lacking.

In general, integrability of a model comes along with an enlargement
of its symmetry structure,
and the establishment of a sufficiently large amount of symmetry
can serve as a definition for integrability.
In the case of planar $\superN=4$ SYM and ABJ(M),
the symmetry structure accompanying integrability
appears to be a Yangian algebra.
Indications for this symmetry were found for
the spectral problem \cite{Dolan:2003uh,Dolan:2004ps},
scattering on the world sheet \cite{Beisert:2006fmy},
scattering amplitudes \cite{Drummond:2009fd}
and smooth Wilson loops \cite{Muller:2013rta}.
The Yangian algebra is a quantum algebra
which typically acts in a non-local fashion.
The non-locality primarily concerns colour space
and a non-local symmetry action can come along with complications.
In this case, the Yangian representation is at odds with
certain cyclic properties of the objects on which it acts.
This principal incompatibility was noticed for all of the above observables,
but it also resolved itself by special features of the model and of the observables.

Due to the above non-locality, the Yangian algebra is not a symmetry of Noether type.
Therefore, there is no natural definition for the Yangian algebra
to serve as a symmetry of a model itself.
Nonetheless, one can resort to the conventional notion that
the invariances of the model's action constitute its symmetries.
For the models at hand, the cyclicity of the action's trace in colour space poses an obstacle.
Another complication arises from the non-linear symmetry action on the fields in these field theory models.
The obstacles were overcome by first considering
Yangian covariance of the equations of motion
and then lifting it to Yangian invariance of the model's action \cite{Beisert:2017pnr,Beisert:2018zxs}.
This is the starting point for this article.

\medskip

The goal of the present article is twofold: first, we propose a more solid 
derivation of the Yangian symmetry statement for the action
put forward in \cite{Beisert:2017pnr,Beisert:2018zxs}, and second,
we apply this derivation to a thorough formulation
of a Yangian symmetry statement for the action of twist deformations of these models
\cite{Leigh:1995ep,Lunin:2005jy,Frolov:2005dj,Beisert:2005if}. 

First, to develop a more solid derivation of Yangian symmetry for the action, 
we reformulate the previous derivation of a Yangian symmetry statement in 
terms of variational forms.
In this variational form framework, the model's action functional is a variational zero-form
and the equations of motion are one-forms, namely the variation of the action zero-form.
Thus, the symmetry statement for the equations of motion in \cite{Beisert:2017pnr} 
translates to a symmetry statement of one-forms.
If this one-form is closed and exact by means of a trivial cohomology,
it will properly integrate to a zero-form
taking the role of a symmetry statement for the action.
For the above planar gauge theory models,
the one-form turns out to be exact
and the integrated zero-form indeed coincides with the symmetry statement for the action proposed in \cite{Beisert:2018zxs}
thus providing a stronger justification for it.
In particular, the variational form framework
maintains manifest cyclicity for the relevant objects at all stages
rendering it a versatile approach for Yangian symmetry
in other field theory models and likely towards quantum correlation functions \cite{Beisert:2018ijg}.

Many integrable models permit deformations that preserve the integrable structure.
A large class of such deformations is given by twist deformations \cite{Leigh:1995ep,Lunin:2005jy,Frolov:2005dj,Beisert:2005if}. 
These deformations can be represented by a linear operator, 
called twist operator acting on the model's action \cite{Drinfeld:1989st,ReshetikhinTwist}.
The second goal of the article is to apply the variational form framework 
to the formulation of Yangian symmetry statements in twist-deformed models,
where the twist also deforms the symmetry generators like the model's action. 
Unfortunately, the twist alters the cyclicity properties of an object on which it acts
unless the latter is uncharged under the charges defining the twist.
This complication imperils a straight-forward analysis of Yangian symmetry for twist-deformed models
for which maintaining cyclicity is essential.
In this respect, the variational form 
framework helps to organise the derivation and to clearly distinguish 
between symmetries of the equations of motion and symmetries of the action. 
Applying the form framework, we find that the equations of motion of the 
twist-deformed model enjoy full deformed Yangian symmetry,
as demonstrated in \cite{Garus:2017bgl}.
Conversely, only the uncharged Yangian symmetries of the equations of motion
can be properly integrated to symmetries of the action.
For the case of beta/gamma-deformed planar gauge theory models, 
the symmetries of the action form an infinite-dimensional quantum algebra
which is a subset of the Yangian algebra
but which itself does not have the form of a complete Yangian algebra and coalgebra.

\medskip

This article is organised as follows: In \secref{sec:forms} the variational 
form framework for Yangian symmetry of the action is developed using the 
symmetry statements derived in \cite{Beisert:2018zxs}. In \secref{sec:twist} 
the variational form framework is applied to twist deformed models and to 
analyse the symmetries of twist deformed planar $\superN=4$ SYM theory. In 
\secref{sec:conclusions} we conclude on the results and provide an outlook 
to further applications.

%%%%%%%%%%%%%%%%%%%%%%%%%%%%%%%%%%%%%%%%%%%%%%%%%%%%%%%%%%%%%%%%%%%%%%%%%%%%%%%%
%%%%%%%%%%%%%%%%%%%%%%%%%%%%%%%%%%%%%%%%%%%%%%%%%%%%%%%%%%%%%%%%%%%%%%%%%%%%%%%%
\section{Exactness of Cyclic Yangian Symmetry}
\label{sec:forms}

In this section we will revisit the proposal
towards Yangian symmetry of the action of 
the planar $\superN=4$ supersymmetric Yang--Mills model 
(along with similar planar gauge theory models) \cite{Beisert:2018zxs}
and put it on a more formal and solid foundation. 
Such a clearer starting point will allow us to argue more thoroughly 
towards a Yangian symmetry of the gamma-deformed $\superN=4$ SYM action 
including its restrictions and prerequisites which are interdependent in subtle ways.

%%%%%%%%%%%%%%%%%%%%%%%%%%%%%%%%%%%%%%%%%%%%%%%%%%%%%%%%%%%%%%%%%%%%%%%%%%%%%%%%
\subsection{Summary of Previous Results}

Let us start by reviewing the results on Yangian symmetry of planar $\superN=4$ SYM
established in \cite{Beisert:2017pnr,Beisert:2018zxs}.

%%%%%%%%%%%%%%%%%%%%%%%%%%%%%%%%%%%%%%%%
\paragraph{Integrability and Yangian Algebra.}

Integrability of classical and quantum mechanical models and/or fields 
comes along with the emergence of a large algebra of (hidden) symmetries,
which typically form some (quantum) deformation of an affine Lie algebra.
In the case of integrability for planar $\superN=4$ SYM, 
the relevant algebra is a quantum algebra of Yangian kind.
Here we will briefly introduce the Yangian algebra and some features
of the relevant representations.

A Yangian algebra is a quantum algebra which can be formulated in many ways.
The simplest presentation is given in terms of 
a set of level-zero generators $\gen^a$
and the corresponding level-one generators $\genyang^a$.
The level-zero generators span a finite-dimensional simple Lie algebra,
and the level-one generators span a corresponding space.
The Yangian algebra is a quantum algebra consisting of polynomials in these generators 
modulo some commutation relations (which we will not consider in this article).
In fact, it is a Hopf algebra, whose coalgebra structure
describe how the generators can act on tensor product representations.
A level-zero generator $\gen$ acts as a homogeneous sum over 
the action $\gen_k$ on a single site of the tensor product
\[
\gen = \sum_k \gen_k.
\]
The level-one generators also have contributions of this kind, 
but more importantly, they also act on pairs of sites
\[
\genyang = \sum_k \genyang_k + \sum_{j<k} \gen^{\swdl{1}}_j \gen^{\swdl{2}}_k .
\]
Here, any term containing the pair $\gen^{\swdl{1}}_j$ and $\gen^{\swdl{2}}_k$
implicitly represents a particular sum of level-zero generators acting on sites $j$ and $k$
(Sweedler notation).
For the Yangian algebra it abbreviates the following pairing
using the dual structure constants $f^c_{ab}$
of the Lie algebra at level zero:
\[
\genyang^c\longrightarrow \gen^{\swdl{1}}\otimes\gen^{\swdl{2}}\defeq\sum_{ab} f^c_{ab}\.\gen^{a}\otimes\gen^{b} .
\]
Note, in particular, that the bilocal terms in the level-one action 
reflect the anti-symmetry of the structure constants.
Assuming that the sites of the tensor product form a lattice,
the level-zero action is called local while the level-one action is called bilocal
referring to the locations along the lattice where the generators act.

%%%%%%%%%%%%%%%%%%%%%%%%%%%%%%%%%%%%%%%%
\paragraph{Invariance Statements.}

In \cite{Beisert:2017pnr,Beisert:2018zxs}, 
it was shown that the equations of motion of the model
are on-shell invariant under a Yangian level-one generator:
Suppose $\eomfor{\field}$ describes the equation of motion $\eomfor{\field}=0$ dual to the field $\field$
and $\gen$ and $\genyang$ denote a level-zero and level-one generator of the Yangian algebra.
The equations of motion can be considered invariant in a weak sense because
the identities
\[
\label{eq:weaksym}
\gen\eomfor{\field}\equiv 0,
\eqsep
\genyang\eomfor{\field}\equiv 0
\]
hold for all fields $\field$
where the equivalence indicates that the identity holds on-shell, 
i.e.\ modulo the equations of motion.
Here, the level-zero generator $\gen$ acts on all fields 
of the equation of motion term $\eomfor{\field}$
and the level-one generator $\genyang$ acts on all pairs of fields of it. 
Note that in the planar limit, the equation of motion term $\eomfor{\field}$
is provided as a non-commutative polynomial
in the fields.
\unskip\footnote{Non-commutative polynomials are understood as the elements
of a free non-commutative monoid ring generated by the fields.} 
The level-one generator thus acts bilocally where locality is 
interpreted in the sense of the adjacency relation of fields within the field monomials.
More concretely, the defining bilocal action on the field monomial $\field_1\field_2\ldots\field_n$ within $\eomfor{\field}$
takes the form
\begin{align}
\label{eq:bilocalsketch}
\numberhere
\gen(\field_1\ldots\field_n)
&=
\sum_{j=1}^n
\field_1\ldots (\gen\field_j)\ldots\field_n,
\\
\genyang(\field_1\ldots\field_n)
&=
\sum_{j=1}^n
\field_1\ldots (\genyang\field_j)\ldots\field_n
+
\sum_{j<k=1}^n
\field_1\ldots (\gen^{\swdl{1}}\field_j)\ldots (\gen^{\swdl{2}}\field_k)\ldots\field_n.
\end{align}
Here, $\gen\field$ denotes the local action of a level-zero generator $\gen$ of the Yangian
which is a generator of the (super)conformal symmetry algebra $\alg{psu}(2,2|4)$ of $\superN=4$ SYM,
and $\genyang\field$ denotes the local action of the level-one generator.

The invariance statement was made significantly stronger by specifying 
the precise combination of the resulting equation of motion terms 
on the right-hand side of the equation
\[
\label{eq:eomsym}
\gen\eomfor{\field}=\sum\nolimits_{\field'}\ast\eomfor{\field}',
\eqsep
\genyang\eomfor{\field}=\sum\nolimits_{\field'}\ast\eomfor{\field}'.
\]
These statements describe the Yangian \emph{covariance of the equations of motion}.
More explicitly, the combination `$\ast$' of equation of motion terms $\eomfor{\field}'$ 
on the right-hand side follows 
by applying the symmetry in a well-prescribed fashion 
to the field $\field$ specifying the particular equation of motion.
Such a symmetry statement provides a fully specific identity for all fields $\field$ 
and for all generators $\gen$ and $\genyang$,
and this identity can serve as a constraint 
in classical and quantum field theoretical considerations of the model.

Now symmetries in theoretical physics, particularly in field theory, 
are usually formulated as an \emph{invariance of the action} $\action$ of a model
such as 
\[
\gen\action=0.
\]
This formulation of symmetries is at the heart of Noether's construction of conserved currents and charges, 
and it also lifts to invariances of correlation functions in quantum field theory
known as Ward--Takahashi identities.
For a planar gauge theory, the invariance under a level-one generator should take the form
\[
\label{eq:actsym}
\genyang\action=0.
\]
The principal difficulty in defining how $\genyang$ acts on $\action$
rests in the cyclic nature of the polynomial terms in the action:
The action is a trace of a non-commutative polynomial,
and as such the cyclicity of the trace identifies 
all monomial terms which are related by a cyclic permutation of the fields. 
Such a projection is at odds with the anti-symmetric definition of the bilocal action 
of level-one generators \eqref{eq:bilocalsketch}.
Nevertheless, and provided certain further assumptions hold,
one can make sense of a level-one action on cyclic polynomials \cite{Drummond:2009fd}.
A precise form for the level-one action $\genyang\action$ 
in a classical setting was proposed in \cite{Beisert:2018zxs}.

%%%%%%%%%%%%%%%%%%%%%%%%%%%%%%%%%%%%%%%%
\paragraph{Relation between Statements.}

The approach of \cite{Beisert:2018zxs} towards defining $\genyang\action$ 
was to compose the Yangian covariance of the equations of motion \eqref{eq:eomsym}
into an invariance statement for the action.
Let us sketch the underlying relationship in terms of an (ordinary) symmetry generator $\gen$.
The relation between the action $\action$ and the equation of motion 
$\eomfor{\field}=0$ dual to the field $\field$ is given by 
\[
\eomfor{\field}\defeq\frac{\var\action}{\var\field}\stackrel{!}{=}0.
\]
The symmetry statement for $\gen$ can be formulated in similar terms as
\unskip\footnote{The abbreviated notation is slightly sloppy
as it discards position as well as colour and flavour indices.
These can however be recovered from a proper interpretation 
of $\sum_{\field'}$ as a generalised sum incorporating an integral over position
and sums over colour and flavour indices.}
\[
0\stackrel{!}{=}
\gen\action
=
\sum\nolimits_{\field'}\gen\field'\. \frac{\var\action}{\var\field'}
%=
%\sum\nolimits_{\field'}\gen\field'\. \eomfor{\field}'
.
\]
Varying this identity with respect to the field $\field$,
one obtains the covariance statement for the corresponding equation of motion
\begin{align}
0\stackrel{!}{=}&\frac{\var(\gen\action)}{\var\field} 
=
\sum\nolimits_{\field'}\gen\field' \frac{\var^2\action}{\var\field\.\var\field'}
+\sum\nolimits_{\field'}\frac{\var (\gen\field')}{\var\field}  \frac{\var\action}{\var\field'}
\\
&=
\sum\nolimits_{\field'}\gen\field' \frac{\var\eomfor{\field}}{\var\field'}
+\sum\nolimits_{\field'}\frac{\var (\gen\field')}{\var\field}  \eomfor{\field}'
\\
&=
\gen\eomfor{\field}
+\sum\nolimits_{\field'}\frac{\var (\gen\field')}{\var\field}  \eomfor{\field}'
.
\end{align}
Hence, the coefficient of the equation of motion dual to $\field'$ 
appearing on the right-hand side of the invariance statement for the 
equation of motion dual to $\field$ is given by 
$-\var (\gen\field')/\var\field$.
The covariance statement for the equation of motions 
therefore follows directly by variation of the corresponding invariance statement for the action.

The proposal was to backtrack the above derivation
for the more elaborate case of a level-one Yangian generator $\genyang$:
In order to undo the variation with respect to a field $\field$,
one combines the assumed covariance of the equations of motion 
$\genyang\eomfor{\field}+\ldots=0$,
with the associated field $\field$ and sums over all fields $\field$.
\unskip\footnote{For completeness, one ought to divide by the degree of 
homogeneity of a monomial in the fields, but this step does
not make much of a difference as all components of the polynomial
with homogeneous degree need to be zero on their own.}
The resulting combination mimics the invariance statement for the action,
and it is zero by construction.
For the level-one Yangian generator $\genyang$
one has to consider some additional identities based on level-zero superconformal symmetry
in order to establish compatibility with cyclic permutations. 
This procedure yields an invariance of the action identity $\genyang\action=0$
which holds by means of construction. 
Nevertheless, it is not a trivial statement because it successfully singles 
out such planar gauge theory models which appear to be integrable.
Furthermore, it provides a non-trivial differential identity for the action 
which can also be lifted to correlation functions within a quantum field theory context \cite{Beisert:2018ijg}.

%%%%%%%%%%%%%%%%%%%%%%%%%%%%%%%%%%%%%%%%
\paragraph{Deformation.}

Subsequently, attempts were made towards lifting the Yangian symmetry results 
to gamma-deformed $\superN=4$ SYM. At the level of algebra, 
the deformation corresponds to a Drinfeld--Reshetikhin twist 
which is a consistent deformation of the Hopf algebra structure.
Indeed, invariance of the equations of motion was shown in \cite{Garus:2017bgl}. 
In particular, the complete Yangian algebra was argued to be a symmetry of this kind
even though some part of the level-zero superconformal symmetry is 
known to be broken by the deformation.
Our goal is to lift the notion of Yangian symmetry for gamma-deformed 
$\superN=4$ SYM to the action.
However, it is difficult to recombine previous results
to an invariance of the action statement using the above procedure.
In parts, this is due to the more elaborate non-local structure
of the deformed symmetries which leaves the recombination procedure ambiguous. 
Furthermore, different forms of the symmetry identities 
for the undeformed model can become inequivalent upon deformation,
and they may or may not hold 
depending on the severity of the deformations of a specific generator.

This suggests that the procedure outlined above is insufficient
for further analysis.
In the following, we offer a reinterpretation and reformulation of 
previous results which offers a more suitable and more rigorous starting point.

%%%%%%%%%%%%%%%%%%%%%%%%%%%%%%%%%%%%%%%%%%%%%%%%%%%%%%%%%%%%%%%%%%%%%%%%%%%%%%%%
\subsection{Variational Forms}

As previously mentioned, the principal difficulty in establishing 
Yangian invariance of the action is the intrinsic cyclicity of the action's field polynomial.
In \cite{Beisert:2018ijg} this difficulty was circumnavigated
by starting with the equations of motion where cyclicity does not apply.
Let us now reformulate and complete the above derivation in terms of variational forms.

%%%%%%%%%%%%%%%%%%%%%%%%%%%%%%%%%%%%%%%%
\paragraph{Invariance Statements as Variational Forms.}

The key insight is that a field variation is a type of derivation, 
and undoing a variation therefore corresponds to integration. 
Here, the fields form a multi-dimensional space and integration
is over a one-dimensional submanifold.
Such integrals describe another field polynomial only if the integrand is integrable,
otherwise the result also depends on the choice of submanifold.
Integrability of the integrand is most conveniently formulated in terms of
differential forms and de Rham cohomology for the field variation operator $\var$: 
a differential form is integrable precisely if it is exact.
Furthermore, de Rham cohomology on polynomials is trivial,
hence a form is exact if and only if it is closed.
\unskip\footnote{The ordinary statement applies to polynomials composed
from commuting objects. However, it apparently holds
for non-commutative and cyclic polynomial as well.}

The above derivation is expressed in terms of forms as follows.
The collection of equations of motion is expressed as
\[
\label{eq:alleomform}
0\stackrel{!}{=} 
\var\action 
= \sum\nolimits_{\field} \var\field\.\frac{\var\action}{\var\field}
= \sum\nolimits_{\field} \var\field\.\eomfor{\field}
.
\]
Here, the right-hand side is a one-form in variations
which are spanned by the basis of elementary field one-forms $\var\field$
(corresponding to the coordinate one-forms). 
The statement implies that the coefficient $\eomfor{\field}$
for every elementary one-form $\var\field$ needs to be zero.

Next, consider the invariance statement for the action under the generator $\gen$
\[
0\stackrel{!}{=}
\invact\defeq\gen\action
= \sum\nolimits_{\field} \gen\field\.\frac{\var\action}{\var\field}
= \sum\nolimits_{\field} \gen\field\.\eomfor{\field}.
\]
In order to turn this statement to a covariance of the equations of motion,
we apply the variational operator $\var$
\begin{align}
0\stackrel{!}{=}
\inveom\defeq
\var\invact
&=
 \sum\nolimits_{\field} \var(\gen\field')\.\eomfor{\field'}
+ \sum\nolimits_{\field} \gen\field'\.\var\eomfor{\field'}
\\
&=
 \sum\nolimits_{\field,\field'} \var\field
\brk[s]*{
\frac{\var(\gen\field')}{\var\field}\eomfor{\field'}
+ \gen\field'\.\frac{\var\eomfor{\field'}}{\var\field}
}
\\
&=
 \sum\nolimits_{\field,\field'} \var\field
\brk[s]*{
\frac{\var(\gen\field')}{\var\field}\eomfor{\field'}
+ \gen\field'\.\frac{\var\eomfor{\field}}{\var\field'}
}
\\
&=
\sum\nolimits_{\field'} \var(\gen\field')\.\eomfor{\field'}
+\sum\nolimits_{\field} \var\field\.\gen\eomfor{\field}
.
\end{align}
Alternatively, we may apply the generator $\gen$ directly to the collective equation of motion
\eqref{eq:alleomform}
\[
0\stackrel{!}{=} 
\inveom'
\defeq\gen(\var\action)
= \sum\nolimits_{\field} \gen(\var\field)\.\eomfor{\field}
+ \sum\nolimits_{\field} \var\field\.\gen\eomfor{\field}
.
\]
Here, we have to specify the level-zero symmetry transformation of a variation $\gen(\var\field)$.
It is reasonable to define it as the variation of the transformed field
\[
\gen(\var\field) \defeq\var (\gen\field).
\]
so that the two operators $\var$ and $\gen$ commute in general
\[
\comm{\var}{\gen}=0.
\]
Both expressions $\inveom$ and $\inveom'$ thus agree.

Let us now consider the opposite approach starting
from covariance of the equations of motion $\inveom=0$:
Here, the one-form $\inveom=\var\invact$ takes on the concrete polynomial expression 
\[
\label{eq:Def_Y_Level0_Invariance}
\inveom=
\sum\nolimits_{\field} \var(\gen\field)\.\eomfor{\field}
+ \sum\nolimits_{\field} \var\field\.\gen\eomfor{\field}.
\]
By construction, this one-form is exact and its integral trivially 
yields the original polynomial $\invact$
expressing invariance of the action $\invact=0$.
However, in ignorance of the (exactness) statement $\inveom=\var\invact$,
a further essential step is to establish that the integrand one-form $\inveom$
is closed and thus integrable. Closedness can readily be confirmed as
\begin{align}
\var \inveom &= 
\sum\nolimits_{\field} \var^2(\gen\field)\.\eomfor{\field}
- \sum\nolimits_{\field} \var(\gen\field)\wedge\var\eomfor{\field}
+ \sum\nolimits_{\field} \var^2\field\.\gen\eomfor{\field}
- \sum\nolimits_{\field} \var\field\wedge\var (\gen\eomfor{\field})
\\
&= 
- \sum\nolimits_{\field,\field'} \var(\gen\field)\wedge\var\field'\frac{\var^2\action}{\var\field\.\var\field'}
- \sum\nolimits_{\field,\field'} \var\field\wedge\var \brk*{\gen\field' \frac{\var^2\action}{\var\field\.\var\field'}}
\\
&= 
- \sum\nolimits_{\field,\field'} \var(\gen\field)\wedge\var\field'\frac{\var^2\action}{\var\field\.\var\field'}
- \sum\nolimits_{\field,\field'} \var\field\wedge\var (\gen\field') \frac{\var^2\action}{\var\field\.\var\field'}
\\ &\alignrel
- \sum\nolimits_{\field,\field',\field''} \var\field\wedge \var\field''
\frac{\var^3\action}{\var\field\.\var\field'\.\var\field''} \gen\field'
\\
&=0.
\end{align}
Here we have made use of the involution property $\var^2=0$ as well as the anti-symmetry
of the wedge product and the symmetry of multiple variations.

It is important to point out that the invariance of the equations of motion $\inveom=0$
of a particular model must not be used towards determining closedness of $\inveom$. 
This would immediately lead to the conclusion $\var \inveom=0$
which would correctly imply that $\inveom$ can be integrated to a constant
and that thus the action would be invariant. 
However, this line of thought is evidently content-free;
effectively, it suffers from an ambiguous definition 
for the invariance statement for the action
from which no conclusion should be drawn. 
Hence, closedness of some expression $\inveom$ must be shown abstractly.

The above reformulation and extension is beneficial for the consideration 
of Yangian symmetry of the action. 
First, it provides a clear-cut criterion for whether a covariance
of the equations of motion actually can be lifted to an invariance of the action 
and thus as a proper symmetry of the model. This criterion is simply 
whether the variational one-form is closed and thus exact.
Below, we shall consider whether it actually holds for the level-one Yangian generators $\genyang$.
Furthermore, it allows us to phrase all derivations and results concerning the 
action and the equations of motion fully in terms of cyclic polynomials.
This will streamline the calculations and offer less distraction from
considering invariance statements for non-cyclic polynomials 
such as the equations of motion.

%%%%%%%%%%%%%%%%%%%%%%%%%%%%%%%%%%%%%%%%
\paragraph{Field Polynomial Notation.}

Finally, let us recall and adjust the notation used in \cite{Beisert:2018ijg}
to represent the planar action of operators on field polynomials.
Here, a field monomial $\oper{O}_{\nfield{n}}$ represents the matrix product of some fields
$\field_1\ldots\field_n$ which are all matrices in colour space,
and a polynomial is a linear combination of such monomials
(with potentially non-homogeneous degree $n$).
In this notation, $\gen_{\nfield{m},k} \oper{O}_{\nfield{n}}$ describes an operator $\gen$
acting on the field $\field_k$ at some site $k=1,\ldots,n$ of the monomial $\oper{O}_{\nfield{n}}$. 
Furthermore, the generator replaces the single field $\field_k$ by 
a field polynomial $\gen_{\nfield{m}}\field_k$ of homogeneous degree $m+1$, 
i.e.\ it effectively adds $m$ new sites to $\oper{O}_{\nfield{n}}$
ranging from site $k$ through site $k+m$.
More explicitly, we can express the result as
\[
\gen_{\nfield{m},k}(\field_1\ldots\field_n)
=
\field_1\ldots\field_{k-1}(\gen_{\nfield{m}}\field_k)\field_{k+1}\ldots\field_n.
\]
This action naturally extends from monomials to polynomials by linearity.

In order to accommodate gauge-invariant objects such as the action $\action$ in the notation, 
we also need to implement traces of products of fields such as $\tr\field_1\ldots\field_n$.
We will represent traced products of fields using the same class of field polynomials,
however, we have to take into account the resulting cyclicity due to the trace in colour space. 
To that end, we introduce the operator $\shift$ 
to represent a cyclic shift of a field polynomial
moving site $1$ to site $2$, i.e.
\[
\shift(\field_1\ldots\field_n)
=
\field_2\ldots\field_n\field_1.
\]
A field polynomial $\oper{O}$ is cyclic if and only if $\shift\oper{O}=\oper{O}$. 
A generic field polynomial $\oper{O}$ can thus represent a product of fields 
as well as its trace in colour space. Traced products of fields, however,
require the representing field polynomial $\oper{O}$ to be cyclic $\shift\oper{O}=\oper{O}$.

As an illustration, we explicitly translate 
the level-zero covariance statement $\inveom_1=0$ for the equations of motion from \cite{Beisert:2018ijg}.
It is written such that the first site of the non-cyclic polynomial $\inveom_1$ enumerates all fields 
and the remaining sites contain the invariance polynomial for the corresponding equation of motion
\[
\label{eq:Level0StrongInvariance}
\inveom_1=
\sum\nolimits_{\field} \field
\brk[s]*{\gen\eomfor{\field}
+\sum\nolimits_{\field'}\frac{\var(\gen\field')}{\var\field}\eomfor{\field}'}.
\]
As such, $\inveom_1=0$ encodes all invariance statements collectively.
We can expand the invariance statement as
\[
\label{eq:Def_Y1_StrongInvariance}
\inveom_1
=
\sum_n\sum_{m=0}^{n-2}\sum_{j=2}^{n-m}\gen_{\nfield{m},j}\action_{\nfield{n-m}}
+
\sum_n\sum_{m=0}^{n-2}\sum_{j=1}^{m+1}\shift^{j-1}\gen_{\nfield{m},1}\action_{\nfield{n-m}},
\]
where the action is expanded as
\[
\action=\sum_n \frac{1}{n}\action_{\nfield{n}}.
\]
In the variational form framework, the field $\field_1$ dual to the equation of motion $\eomfor{\field}_1$ 
is marked as a field variation $\var\field_1$. 
Within $\inveom_1$ we can mark it as $\var_1\inveom_1$,
where we have used the notation $\var_k$ 
to indicate the action of the variational operator $\var$ on site $k$.
With the first field marked explicitly, 
we are now free to convert the (not manifestly cyclic) polynomial $\inveom_1$
to a manifestly cyclic polynomial one-form $\inveom$
\[
\inveom
=\tr\var_1\inveom_1,
\]
where we have defined the trace operator `$\tr$' as the cyclic average operator
\unskip\footnote{\label{fn:Def_tr}%
A trace of matrix-valued fields is cyclic,
therefore closing a non-cyclic polynomial by means of a trace
effectively projects it to its cyclic part.
The notation `$\tr$' is thus understood as a linear operator on field polynomials
rather than an action on matrices in colour space.}
\[
\tr \oper{O}_{\nfield{n}}\defeq\frac{1}{n}\sum_{j=1}^n \shift^j\oper{O}_{\nfield{n}}.
\]
The cyclic average is a projection operator, and on $\oper{O}_{\nfield{n}}=\field_1\ldots\field_n$
it acts as
\[
\label{eq:Def_tr}
\tr(\field_1\ldots\field_n)= 
\frac{1}{n}
\sum_{j=1}^n
\field_j\ldots\field_n \field_1\ldots\field_{j-1}.
\]
The cyclic projection is helpful for our purposes 
as we can remain within the space of cyclic polynomials,
i.e.\ both $\action$ and $\inveom$ are manifestly cyclic.
The trace allows us to shift around some operators by means of powers of $\shift$ 
and we simplify
\begin{align}
\label{eq:InvEom_0}
\inveom
&= 
\sum_n\sum_{m=0}^{n-2} \sum_{j=2}^{n-m} \tr\var_{n+2-j}\gen_{\nfield{m},1}\action_{\nfield{n-m}}
+\sum_n\sum_{m=0}^{n-2} \sum_{j=1}^{m+1} \tr\var_j\gen_{\nfield{m},1}\action_{\nfield{n-m}}
\\
&= 
\sum_n\sum_{m=0}^{n-2} \sum_{j=1}^{n} \tr\var_j\gen_{\nfield{m},1}\action_{\nfield{n-m}}
=
\sum_n\sum_{m=0}^{n-2} \tr \var(\gen_{\nfield{m},1}\action_{\nfield{n-m}})
\\
&=
\sum_n\sum_{m=0}^{n-2} \sum_{j=1}^{n-m}\frac{1}{n-m}\tr \var(\gen_{\nfield{m},j}\action_{\nfield{n-m}})
=\tr \var (\gen\action).
\end{align}
This one-form $\inveom$ is manifestly exact and closed,
and its form agrees with the above invariance statement.
We can now integrate the one-form $\inveom$ to the zero-form
\[
\invact=\tr\gen\action,
\]
and $\invact=0$ is indeed the proper invariance statement for the action.

Note that the above derivation of exactness of $\inveom$ could be streamlined
by making use of the de-facto cyclicity of the expression $\inveom_1$:
\[
\label{eq:closedbycyclicity}
\inveom_{\nfield{n}}
=\tr\var_1\inveom_{1,\nfield{n}} 
=\frac{1}{n}\sum_{j=1}^n \tr\var_1\shift^{j-1} \inveom_{1,\nfield{n}} 
=\frac{1}{n}\sum_{j=1}^n \tr\var_j \inveom_{1,\nfield{n}} 
= \frac{1}{n} \var \tr\inveom_{1,\nfield{n}}.
\]
Here, we have split up the statement into components of uniform homogeneity $n$.
Then the factor $1/n$ effectively implements the integration of a polynomial term
of this degree.

%%%%%%%%%%%%%%%%%%%%%%%%%%%%%%%%%%%%%%%%%%%%%%%%%%%%%%%%%%%%%%%%%%%%%%%%%%%%%%%%
\subsection{Level-One Symmetry using Variational Forms}

We can now address the level-one bilocal symmetries using variational forms.

%%%%%%%%%%%%%%%%%%%%%%%%%%%%%%%%%%%%%%%%
\paragraph{Equations of Motion.}

The level-one invariance of the equations of motion $\inveomyang_1=0$ 
was expressed in \cite{Beisert:2018ijg} in terms of the non-cyclic polynomial
\[
\label{eq:DefYchcC}
\inveomyang_1 
=
\sum_{\field}
\field
\brk[s]*{\genyang\frac{\var \action}{\var \field}
+ 
\sum_{\field'}
\frac{\var \action}{\var \field'} \frac{\var(\genyang\field')}{\var \field} 
- \sum_{\field'}\frac{\var \action}{\var \field'} \brk*{\gen^{\swdl{1}} \wedge \frac{\var}{\var \field}}
  \brk{\gen^{\swdl{2}}\field'}  }.
\]
Here, the wedge product term in the brackets 
represents a certain anti-symmetric action on the 
non-linear output of $\gen^{\swdl{2}}\field'$,
see \cite{Beisert:2018ijg} for details.
In the field polynomial notation, the expression translates to
\begin{align}
\label{eq:Def_Y1h_StrongInvariance}
\inveomyang_1
&=
\sum_n
\sum_{m=0}^{n-2}
\sum_{j=2}^{n-m}
\genyang_{\nfield{m},j}
\action_{\nfield{n-m}}
+
\sum_n
\sum_{m=0}^{n-2}
\sum_{j=1}^{m+1}
\shift^{j-1}
\genyang_{\nfield{m},1}
\action_{\nfield{n-m}}
\\
&\alignrel{}
+
\sum_n
\sum_{m=0}^{n-2}
\sum_{l=0}^{m}
\sum_{j=2}^{n-m-1}\sum_{k=j+1}^{n-m}
\gen^{\swdl{1}}_{\nfield{l},j}
\gen^{\swdl{2}}_{\nfield{m-l},k}
\action_{\nfield{n-m}}
\\
&\alignrel{}
+
\sum_n
\sum_{m=0}^{n-2}
\sum_{l=0}^{m}
%\sum_{j=1}^{l}\sum_{k=j+1}^{l+1}
%\brk!{
%\shift^{j-1}\gen^{\swdl{1}}_{\nfield{m-l},k}
%-\shift^{k-1+m-l}\gen^{\swdl{1}}_{\nfield{m-l},j}
%}
%\gen^{\swdl{2}}_{\nfield{l},1}
%\action_{\nfield{n-m}}.
\sum_{k=1}^{l+1}
\brk[s]*{\sum_{j=1}^{k-1}-\sum_{j=k+m-l+1}^{m+1}}
\shift^{j-1}
\gen^{\swdl{1}}_{\nfield{m-l},k}
\gen^{\swdl{2}}_{\nfield{l},1}
\action_{\nfield{n-m}}.
\end{align}
The first site of the polynomial $\inveomyang_1$ specifies a field
dual to an equation of motion and the remaining sites yield the 
invariance statement for this equation of motion. 
Consequently, $\inveomyang_1=0$ summarises the invariance statements
for all equations of motion.
In our framework, we express this term as a cyclic one-form
\[
\label{eq:Def_Y0_StrongInvarianceCyc}
\inveomyang_{\textnormal{eom}}
\defeq
\tr\var_1\inveomyang_1.
\]
%
%\begin{align}
%\inveomyang_{\textnormal{eom}}
%\defeq
%\tr\var_1\inveomyang_1
%&=
%\sum_n
%\sum_{m=0}^{n-2}
%\sum_{j=1}^{n}
%\tr\var_j
%\genyang_{\nfield{m},1}
%\action_{\nfield{n-m}}
%\\
%&\alignrel{}
%+
%\sum_n
%\sum_{m=0}^{n-2}
%\sum_{l=0}^{m}
%\sum_{j=2}^{n-m-1}\sum_{k=j+1}^{n-m}
%\tr\var_1
%\gen^{\swdl{1}}_{\nfield{l},j}
%\gen^{\swdl{2}}_{\nfield{m-l},k}
%\action_{\nfield{n-m}}
%\\
%&\alignrel{}
%+
%\sum_n
%\sum_{m=0}^{n-2}
%\sum_{l=0}^{m}
%\sum_{j=1}^{l}\sum_{k=j+1}^{l+1}
%\tr
%\brk!{
%\gen^{\swdl{1}}_{\nfield{m-l},k}\var_j
%-\gen^{\swdl{1}}_{\nfield{m-l},j}\var_k
%}
%\gen^{\swdl{2}}_{\nfield{l},1}
%\action_{\nfield{n-m}}
%\end{align}
%
Here the field variation $\var \field$ specifies the field dual to the equation of motion.
As this element can now reside at any place within the polynomial,
it makes sense to project to cyclic polynomials.
The vanishing of the cyclic polynomial $\inveomyang_{\textnormal{eom}}$ now expresses invariance of the equations of motion.

%%%%%%%%%%%%%%%%%%%%%%%%%%%%%%%%%%%%%%%%
\paragraph{Closedness.}

In \cite{Beisert:2018ijg} it was shown that $\inveomyang_1$ is effectively cyclic, 
and the cyclic average was assumed to be the symmetry variation of the action
(after a tacit division by the degree of the monomial terms).
This provided an invariance statement for the action which worked adequately.
\unskip\footnote{This reasoning can be justified
along the lines of the argument around \eqref{eq:closedbycyclicity}.}
It however remained somewhat unclear why the procedure worked and to what extent 
the invariance statement is unique or appropriate. 
For instance, it is not at all evident why one should take a cyclic average
and not some other procedure to extract a cyclic term from a non-cyclic one. 
Such a question is particularly relevant for gamma-deformed planar gauge theories
where the right procedure is not evident.
The issue is resolved by using variational forms where there is a well-defined and consistent procedure
as outlined above.

Our aim is to integrate the one-form $\inveomyang_{\textnormal{eom}}$ to a zero-form $\invactyang$.
In other words, the one-form $\inveomyang_{\textnormal{eom}}$ must be exact, $\inveomyang_{\textnormal{eom}}=\var\invactyang$, 
and a prerequisite as well as sufficient condition is that
$\inveomyang_{\textnormal{eom}}$ is closed, $\var\inveomyang_{\textnormal{eom}}=0$.
It is straight-forward to compute the second variation
\begin{align}
\numberhere
\var\inveomyang_{\textnormal{eom}}
&=
\sum_n
\sum_{m=0}^{n-2}
\sum_{l=0}^{m}
\sum_{i=1}^n
\sum_{j=2}^{n-m-1}
\sum_{k=j+1}^{n-m}
\tr\var_i\var_1
\gen^{\swdl{1}}_{\nfield{l},j}
\gen^{\swdl{2}}_{\nfield{m-l},k}
\action_{\nfield{n-m}}
\\
&\alignrel{}
+
\sum_n
\sum_{m=0}^{n-2}
\sum_{l=0}^{m}
\sum_{i=1}^n
\sum_{j=1}^{l}
\sum_{k=j+1}^{l+1}
\tr
\brk!{
\var_i\var_j\gen^{\swdl{1}}_{\nfield{m-l},k}
-\var_i\var_{k+m-l}\gen^{\swdl{1}}_{\nfield{m-l},j}
}
\gen^{\swdl{2}}_{\nfield{l},1}
\action_{\nfield{n-m}}.
\end{align}
These remaining terms do not cancel, but in fact they can be rearranged as
\begin{align}
\var\inveomyang_{\textnormal{eom}}
&=
\sum_n
\sum_{m=0}^{n-2}
\sum_{j=m+2}^{n-1}
\sum_{k=j+1}^{n}
\tr\var_j\var_k
\brk!{
\gen^{\swdl{1}}_{\nfield{m},1}
\invact^{\swdl{2}}_{\nfield{n-m}}
-\gen[H]_{\nfield{m},1}
\action_{\nfield{n-m}}
}
\\
&\alignrel{}
+
\sum_n
\sum_{m=0}^{n-2}
\sum_{j=1}^{m}
\sum_{k=j+1}^{m+1}
\tr\var_j\var_k
\brk!{
\gen^{\swdl{1}}_{\nfield{m},1}
\invact^{\swdl{2}}_{\nfield{n-m}}
-\gen[H]_{\nfield{m},1}
\action_{\nfield{n-m}}
}.
\end{align}
Here, $\invact^{\swdl{i}}$ represents the level-zero symmetry variation of the action under the generator $\gen^{\swdl{i}}$
\[
\invact^{\swdl{i}}\defeq
\tr\gen^{\swdl{i}}\action=
\sum_n
\frac{1}{n}\invact^{\swdl{i}}_{\nfield{n}},
\eqsep
\invact^{\swdl{i}}_{\nfield{n}}
= 
\sum_{m=0}^{n-2}\sum_{j=1}^n \shift^{j-1} \gen^{\swdl{i}}_{\nfield{m},1}\action_{\nfield{n-m}},
\]
and $\gen[H]\defeq\half\comm{\gen^{\swdl{1}}}{\gen^{\swdl{2}}}$
represents the commutator term producing the dual Coxeter number
for a simple Lie algebra
\[
\gen[H]_{k}=\sum_n \gen[H]_{\nfield{n},k},
\eqsep
\gen[H]_{\nfield{n},k}=
\sum_{m=0}^n\sum_{j=0}^m
\gen^{\swdl{1}}_{\nfield{n-m},k+j}\gen^{\swdl{2}}_{\nfield{m},k}.
\]
It is a prerequisite for Yangian symmetry of a planar gauge model 
that both of these quantities are zero.
Hence, the invariance one-form $\inveomyang_{\textnormal{eom}}$ is effectively closed 
and could be integrated in principle.

Nevertheless, we first have to find a manifestly exact expression:
Closedness follows from specific features of the model,
and in order to achieve it, we have to work in the specific model. 
In that case, however, the invariance of the equations of motion
makes the one-form $\inveomyang_{\textnormal{eom}}$ vanish,
and as a consequence, closedness and exactness of $\inveomyang_{\textnormal{eom}}$ become trivial statements.
Eventually, also the integral $\invactyang$ is zero by construction, but one cannot draw 
a meaningful conclusion regarding invariance of the action from this.

Therefore, we should first find an expression
\[
\inveomyang=\inveomyang_{\textnormal{eom}}+\inveomyang_{\mathrm{\Delta}},
\]
which is exactly closed irrespective of the model at hand
and where the correction terms $\inveomyang_{\mathrm{\Delta}}$
are proportional to $\invact$ or $\gen[H]$. 
Consequently, $\inveomyang_{\mathrm{\Delta}}=0$ for models of interest, 
and $\inveomyang$ is equivalent to $\inveomyang_{\textnormal{eom}}$ for all purposes. 
Importantly, closedness of $\inveomyang$ now allows us to integrate it irrespectively of the model,
\[
\var\inveomyang=0
\eqjoin{\implies}
\inveomyang=\var\invactyang.
\]
Closedness implies that the term $\inveomyang_{\mathrm\Delta}$ 
takes the almost unique form:
\begin{align}
\label{eq:CorrectionL1}
\numberhere
\inveomyang_{\mathrm\Delta}
&=
-\sum_n\sum_{m=0}^{n-2}
\tr 
\brk!{\gen^{\swdl{1}}_{\nfield{m},1}\var_1\invact^{\swdl{2}}_{\nfield{n-m}}
  -\gen[H]_{\nfield{m},1}\var_1\action_{\nfield{n-m}}}
\\
&\alignrel
+\sum_n\sum_{m=0}^{n-2}
\sum_{j=1}^{n}
\frac{m+2-2j}{n}
\tr \var_{j}
\brk!{\gen^{\swdl{1}}_{\nfield{m},1}\invact^{\swdl{2}}_{\nfield{n-m}}
  -\gen[H]_{\nfield{m},1}\action_{\nfield{n-m}}}.
\end{align}
We note that the coefficients in the second line
could be adjusted slightly to produce additional exact contributions
thanks to the identity
\[
\sum_{m=0}^{n-2}
\tr\gen^{\swdl{1}}_{\nfield{m},1}
\invact^{\swdl{2}}_{\nfield{n-m}}
=
\sum_{m=0}^{n-2}
\tr
\gen[H]_{\nfield{m},1}
\action_{\nfield{n-m}}.
\]
The form chosen above leads to a result with a reasonably symmetric structure.

%%%%%%%%%%%%%%%%%%%%%%%%%%%%%%%%%%%%%%%%
\paragraph{Integration.}

We can now perform the integral of the one-form $\inveomyang$.
This yields precisely the level-one symmetry variation of the action 
$\invactyang$ as proposed in \cite{Beisert:2018ijg}
\begin{align}
\label{eq:Xhat}
\numberhere
\invactyang
&=
\sum_n\sum_{m=0}^{n-2}
\tr\genyang_{\nfield{m},1}
\action_{\nfield{n-m}}
\\
&\alignrel{}
+\sum_n\sum_{m=0}^{n-2}
\sum_{l=0}^{m}
\sum_{k=1}^{n-m+2l+1} \frac{k-l-\half n+\half m-1}{n}
\tr\gen^{\swdl{1}}_{\nfield{m-l},k}
\gen^{\swdl{2}}_{\nfield{l},1}
\action_{\nfield{n-m}}
%\\
%&\alignrel{}
%+\sum_n\sum_{m=0}^{n-2}
%c_{n,m}
%\tr\gen^{\swdl{1}}_{\nfield{m},1}
%\invact^{\swdl{2}}_{\nfield{n-m}}
\\
&=
\sum_n\sum_{m=0}^{n-2}
\tr\genyang_{\nfield{m},1}
\action_{\nfield{n-m}}
\\
&\alignrel{}
+\sum_n\sum_{m=0}^{n-2}
\sum_{l=0}^{m}
\sum_{k=2}^{n-m} \frac{k-\half n+\half m-1}{n}
\tr\gen^{\swdl{1}}_{\nfield{m-l},k+l}
\gen^{\swdl{2}}_{\nfield{l},1}
\action_{\nfield{n-m}}
\\
&\alignrel{}
+\sum_n\sum_{m=0}^{n-2}
\sum_{l=0}^{m}
\sum_{k=1}^{l+1} \frac{2k-l-2}{n}
\tr\gen^{\swdl{1}}_{\nfield{m-l},k}
\gen^{\swdl{2}}_{\nfield{l},1}
\action_{\nfield{n-m}}
.
\end{align}
More accurately, we extracted the above additional terms 
$\inveomyang_{\mathrm\Delta}\defeq\var\invactyang-\inveomyang_{\textnormal{eom}}$
by comparing with the expected integral $\invactyang$ from \cite{Beisert:2018ijg}.
Supposing that the previously provided $\invactyang$ is the correct symmetry variation of the action, 
this procedure will yield a consistent expression $\inveomyang_{\mathrm\Delta}$.
Here, $\inveomyang_{\mathrm\Delta}$ indeed leads to an exactly closed $\inveomyang$,
which justifies the correctness of the expression $\invactyang$.

%%%%%%%%%%%%%%%%%%%%%%%%%%%%%%%%%%%%%%%%%%%%%%%%%%%%%%%%%%%%%%%%%%%%%%%%%%%%%%%%
%%%%%%%%%%%%%%%%%%%%%%%%%%%%%%%%%%%%%%%%%%%%%%%%%%%%%%%%%%%%%%%%%%%%%%%%%%%%%%%%
\section{Twist Deformation}
\label{sec:twist}

We are now in a good position to address Yangian symmetry of the twist-deformed model.
The twist deformation can be expressed as a transformation on field polynomials
which takes the undeformed model to the deformed one.
On the one hand, this transformation extends directly to the action of Yangian generators
and thus to symmetry statements. 
On the other hand, it breaks cyclicity of certain field polynomials under some well-prescribed conditions.
Here, the formalism introduced in \secref{sec:forms} allows us 
to unambiguously express symmetry statements 
and to analyse how far the latter remain compatible with cyclicity.
In doing so, we will observe that the twist deformation applies in a different manner
to the non-cyclic covariance of the equations of motion and to invariance of the cyclic action.
We thus find that the twist deformation reduces the Yangian symmetry of planar $\superN=4$ SYM
to a particular subalgebra.

%%%%%%%%%%%%%%%%%%%%%%%%%%%%%%%%%%%%%%%%%%%%%%%%%%%%%%%%%%%%%%%%%%%%%%%%%%%%%%%%
\subsection{Deformed Model}

We start by introducing the beta/gamma-deformed $\superN=4$ SYM model 
in terms of a Drinfeld--Reshetikhin twist transformation of the undeformed model.
The Drinfeld--Reshetikhin twist \cite{Drinfeld:1989st,ReshetikhinTwist}
is a consistent deformation of a quantum algebra in terms of its Cartan subalgebra. 
In the case of the $\superN=4$ SYM model, 
we restrict our attention to the Cartan subalgebra of the internal $\alg{su}(4)$ symmetry
and hence there are three relevant Cartan generators $\gencart^a$ with $a=1,2,3$.
Further, let $\gamma_{ab}$ denote a real, anti-symmetric $3\times 3$ matrix. 
For generic $\gamma_{ab}$ with three independent degrees of freedom,
the deformed model is called the gamma-deformation,
whereas a one-parameter subspace of matrices $\gamma_{ab}$
preserving $\superN=1$ supersymmetry 
is known as the beta-deformation \cite{Leigh:1995ep,Lunin:2005jy,Frolov:2005dj}. 

%%%%%%%%%%%%%%%%%%%%%%%%%%%%%%%%%%%%%%%%
\paragraph{Drinfeld--Reshetikhin Twist.}

The Drinfeld--Reshtikhin twist 
\cite{Drinfeld:1989st,ReshetikhinTwist}
is defined by a two-site operator 
\[
\twist \defeq \exp\brk*{\iunit\gamma_{ab}\gencart^a\otimes\gencart^b }.
\]
Many two-site structures of the Hopf algebra underlying integrability
such as the coproduct governing the two-site action of $\genyang$
are deformed by multiplication or conjugation with the twist operator $\twist$.
This operator extends to an arbitrary number of sites 
and thus to field polynomials as
\[
\label{eq:DefTwist}
\twist \oper{O}_{\nfield{n}}  = 
\prod_{j=1}^{n-1}\prod_{k=j+1}^n
\exp \brk!{\iunit\gamma_{ab}\gencart^a_j\otimes\gencart^b_k }
\oper{O}_{\nfield{n}}. 
\]
In particular, the deformed model is obtained from the undeformed model 
by twisting the action \cite{Lunin:2005jy}
\[
\action^\twiststar \defeq \twist\action.
\]
Here and in the following, we decorate quantities 
of the twisted model by a star, e.g.\ the twisted action is denoted by $\action^\twiststar$.

%%%%%%%%%%%%%%%%%%%%%%%%%%%%%%%%%%%%%%%%
\paragraph{Compatibility with Cyclicity.}

A distinctive feature of the twist transformation $\twist$ in \eqref{eq:DefTwist} 
is that it assumes a fixed ordering of the sites with $1\leq j<k\leq n$.
In that form, it is not a priori well-defined on cyclic polynomials such as the action polynomial $\action$ of $\superN=4$ SYM. 
More concretely, the twist transformation does not commute with the
cyclic shift operation
\[
\label{eq:ShiftAndTwist}
 \twist \shift \neq \shift \twist.
\]
For a cyclic polynomial $\oper{O}=\shift\oper{O}$ 
this implies that the twisted image $\oper{O}^\twiststar\defeq\twist\oper{O}$ 
is not generically cyclic itself: 
$\oper{O}^\twiststar=\twist\oper{O}=\twist\shift\oper{O}\neq \shift\twist\oper{O}=\shift\oper{O}^\twiststar$.
Nevertheless, cyclicity does hold for the deformed action $\action^\twiststar$
because the action is invariant under the Cartan generators $\gencart^a$ defining the twist,
namely $\gencart^a\action=0$.
More generally, the twisted image $\oper{O}^\twiststar$ of a cyclic polynomial $\oper{O}$ is 
itself cyclic if and only if it is uncharged with respect to the twist deformation
as follows
\[
\gamma_{ab}\gencart^b \oper{O}=0.
\]

It is instructive to review the statement explicitly:
Let $\oper{O}_1\defeq AB$ where $A$ and $B$ can be two individual fields or some other sub-monomials.
Define further $\oper{O}_2\defeq BA=\shift^k\oper{O}_1$ as some cyclic image of $\oper{O}_1$
($k$ equals the number of fields in $A$).
The twisted images $\oper{O}_{1,2}^\twiststar\defeq\twist \oper{O}_{1,2}$ of the monomials read
\[
\oper{O}_1^\twiststar
=
\exp \brk!{\iunit\gamma_{ab}\.\chargecart^a[A]\.\chargecart^b[B] }
A^\twiststar B^\twiststar ,
\eqsep
\oper{O}_2^\twiststar
=
\exp \brk!{\iunit\gamma_{ab}\.\chargecart^a[B]\.\chargecart^b[A] }
B^\twiststar A^\twiststar .
\]
Here, $\chargecart^a[X]$ denotes the eigenvalue of $\gencart^a$ acting on the field expression $X$
(supposing that $X$ is indeed an eigenvector of $\gencart^a$). 
Now, the monomial $\oper{O}_2^\twiststar\sim B^\twiststar A^\twiststar $ 
is still a cyclic image of $\oper{O}_1^\twiststar\sim A^\twiststar B^\twiststar$,
yet with a different prefactor.
When this statement is applied to the monomials constituting a cyclic polynomial, 
different prefactors apply to different monomials and
it is evident that the twisted image is typically not cyclic.
However, supposing that the monomials $\oper{O}_{1,2}$ are altogether uncharged with respect to the twist, 
we have that
\[
\gamma_{ab}\.\chargecart^b[\oper{O}_{1,2}]=\gamma_{ab}\.\chargecart^b[A]+\gamma_{ab}\.\chargecart^b[B]=0
\eqjoin{\implies}
\gamma_{ab}\.\chargecart^b[B]=-\gamma_{ab}\.\chargecart^b[A].
\]
In that case, both exponents reduce to $-\iunit\gamma_{ab}\chargecart^a[A]\chargecart^b[A]$
which is zero due to anti-symmetry of the twist matrix $\gamma_{ab}$.
Then, $\oper{O}_2^\twiststar=\shift^k \oper{O}_1^\twiststar$ is a cyclic image of $\oper{O}_1^\twiststar$ with factor $1$.
As this statement is independent of the concrete splitting of some monomial $\oper{O}_1=AB$ into sub-monomials $A,B$,
it extends to generic polynomials. 
In other words, as far as the monomials contributing to a polynomial are properly mapped to each other 
such that the resulting polynomial is cyclic, the same will hold for the twisted images.

%%%%%%%%%%%%%%%%%%%%%%%%%%%%%%%%%%%%%%%%%%%%%%%%%%%%%%%%%%%%%%%%%%%%%%%%%%%%%%%%
\subsection{Twisted Algebra}

The twist $\twist$ constitutes a linear transformation,
consequently, other linear operators are deformed 
by conjugation with $\twist$. 
In particular this applies to the twisted action of the Yangian generators
\[
\label{eq:TwistedRepresentation}
\gen^\twiststar\coloneqq \twist \gen \twist^{-1},
\eqsep
\genyang^\twiststar \coloneqq \twist \genyang \twist^{-1}.
\]
The construction via conjugation ensures 
that twisted polynomials transform under the twisted Yangian action
exactly in the same way as the undeformed polynomials transform under the untwisted Yangian action. 

Nevertheless, there are important differences between 
the twisted generators $\gen^\twiststar$ and $\genyang^\twiststar$ 
and their undeformed counterparts $\gen$ and $\genyang$ in terms of locality. 
Let us discuss this aspect in more detail because it is relevant for the action on cyclic polynomials.
The generators $\gen$ act locally with regard to the ordering of fields within field polynomials, 
while the $\genyang$ act bilocally,
see \eqref{eq:bilocalsketch}.
Conversely, the twist transformation $\twist$ in \eqref{eq:DefTwist} is an exponential of bilocal terms, 
and as such it acts rather non-locally.
This non-locality can reduce somewhat when $\twist$ conjugates $\gen$ and $\genyang$.

%%%%%%%%%%%%%%%%%%%%%%%%%%%%%%%%%%%%%%%%
\paragraph{Level-Zero Generators.}

Within the conjugation $\gen^\twiststar=\twist\gen\twist^{-1}$,
one of the two legs of the bilocal term in the exponent of $\twist$ 
must meet with the local term in $\gen$ in order to contribute non-trivially to the twist. 
As such, it effectively measures the Cartan charge $\chargecart^a[\gen]$ of the local term.
Altogether, this leads to the insertion of the group-like operator
\[
\label{eq:gentwist}
\gentwist\defeq \exp(\iunit\gamma_{ab}\.\chargecart^a[\gen]\. \gencart^b),
\eqsep
\gentwist_{j,k}\defeq \gentwist_{j}\cdots \gentwist_{k},
\]
acting as $\gentwist^{-1}$ or $\gentwist$ on all sites
to the left or right, respectively, of the local term $\gen$,
\begin{align}
\label{eq:TwistedRepresentationCharged}
\gen^\twiststar (\field_1\ldots\field_n)
&=
\sum_{k=1}^n 
(\gentwist^{-1}\field_1)\ldots (\gentwist^{-1}\field_{k-1})
\.(\gen^\twiststar\field_k)\.
(\gentwist\field_{k+1})\ldots (\gentwist\field_n)
\\
&=
\sum_{k=1}^n 
\brk!{ \gen^\twiststar_k \.\gentwist^{-1}_{1,k-1}\gentwist_{k+1,n}}
(\field_1\ldots\field_n).
\end{align}
These additional factors clearly distinguish the action of 
charged generators $\gen$ ($\gamma_{ab}\chargecart^a[\gen]\neq0$) 
from the action of uncharged generators $\gen$ ($\gamma_{ab}\chargecart^a[\gen]=0$): 
For the latter, the corresponding group-like transformation $\gentwist$ is the identity map,
and therefore the twisted generator $\gen^\twiststar$ describes a local action 
just as the original generator $\gen$
\[
\label{eq:TwistedRepresentationUncharged}
\gen^\twiststar (\field_1\ldots\field_n) =
\sum_{k=1}^n 
\field_1\ldots \field_{k-1}
\.(\gen^\twiststar\field_k)\.
\field_{k+1}\ldots \field_n.
\]
Conversely, charged twisted generators $\gen^\twiststar$ act non-locally on field polynomials
as in \eqref{eq:TwistedRepresentationCharged}.
In particular, the non-local terms manifestly spoil the cyclic properties of the polynomials on which they act.
This is apparent from the fact that the factors $\gentwist^{-1}$ on the far left 
and the factors $\gentwist$ on the far right are inverses of each other,
whereas the cyclicity assumes these to be nearest neighbours
without anything significant in between.

We may view this issue as a manifestation of the distinct cyclicity features of 
the twist operator $\twist$ acting on charged vs.\ uncharged states discussed 
below \eqref{eq:ShiftAndTwist}:
Let us consider acting with $\gen^\twiststar=\twist\gen\twist^{-1}$ on an uncharged cyclic state $\oper{O}$
step by step.
First, $\twist^{-1}\oper{O}$ is cyclic and uncharged due to the above argument.
If now $\gen$ is uncharged, so will $\gen\twist^{-1}\oper{O}$ be,
and finally, $\gen^\twiststar\oper{O}=\twist\gen\twist^{-1}\oper{O}$ will be cyclic and uncharged.
If, conversely, $\gen$ is charged, $\gen\twist^{-1}\oper{O}$ will be cyclic and charged,
and finally, $\gen^\twiststar\oper{O}=\twist\gen\twist^{-1}\oper{O}$ will be charged and not cyclic.
Therefore, the action of a charged $\gen^\twiststar$ on a cyclic $\oper{O}$ 
maps to a non-cyclic state which does not make physical sense.

%%%%%%%%%%%%%%%%%%%%%%%%%%%%%%%%%%%%%%%%
\paragraph{Level-One Generators.}

Similar restrictions apply to the level-one generators $\genyang$.
We therefore might focus exclusively on uncharged generators $\genyang$ right away, 
but let us remain open-minded for the time being.
The above considerations apply directly to local contributions to $\genyang$.
The bilocal contributions with $\gen^{\swdl{1}}$ acting towards the left of $\gen^{\swdl{2}}$
split up the field polynomial into three regions
where different group-like factors apply as follows:
towards the left of $\gen^{\swdl{1}}$ the relevant factor is given by
$\gentwist^{\swdl{1}\,-1}\gentwist^{\swdl{2}\,-1}=\gentwistyang^{-1}$;
towards the right of $\gen^{\swdl{2}}$
it is $\gentwist^{\swdl{1}}\gentwist^{\swdl{2}}=\gentwistyang$
and between $\gen^{\swdl{1}}$ and $\gen^{\swdl{2}}$ the applicable factor is 
$\gentwist^{\swdl{1}}\gentwist^{\swdl{2}\,-1}=\gentwist^{\swdl{1/2}}$
with the following extensions of the definition \eqref{eq:gentwist}:
\[
\gentwist^{\swdl{i}}\defeq \exp(\iunit\gamma_{ab}\.\chargecart^a[\gen^{\swdl{i}}]\. \gencart^b),
\eqsep
\gentwistyang\defeq \exp(\iunit\gamma_{ab}\.\chargecart^a[\genyang]\. \gencart^b),
\eqsep
\gentwist^{\swdl{i/j}}\defeq\gentwist^{\swdl{i}}\gentwist^{\swdl{j}\,-1}.
\]
In analogy to \eqref{eq:TwistedRepresentation},
the qualitative features of the action could be abbreviated as
\[
\genyang^\twiststar =
\operatorname{local}+
\sum_{j<k} 
\gentwistyang^{-1}_{1,j-1}\.
\gen^{\swdl{1}}_j\.
\gentwist^{\swdl{1}}_{k}
\gentwist^{\swdl{1/2}}_{j+1,k-1}
\gentwist^{\swdl{2}\,-1}_{j}\.
\gen^{\swdl{2}}_k\.
\gentwistyang_{k+1,n}.
\]
Therefore, the applicable factors on the left and right asymptotic regions
are determined by the overall generator $\genyang$ alone, 
while the applicable factor between the two legs
$\gen^{\swdl{1}}$ and $\gen^{\swdl{2}}$ depends on the concrete splitting
of $\genyang$ into $\gen^{\swdl{1}}$ and $\gen^{\swdl{2}}$.
In any case, the applicable factors on the far left and on the far right
are distinct if and only if $\genyang$ is charged.
This feature is in full agreement with the behaviour of the local contribution to $\genyang^\twiststar$. 
Note, however, that the bilocal contribution to $\genyang^\twiststar$ is still substantially non-local even if $\genyang$ is uncharged
due to the non-trivial group-like factors $\gentwist^{\swdl{1/2}}$
acting between the bilocal insertions $\gen^{\swdl{1}}$ and $\gen^{\swdl{2}}$.
On the one hand, these factors are well-separated by the insertions $\gen^{\swdl{1}}$ and $\gen^{\swdl{2}}$
and thus do not represent an obstacle towards proper cyclicity
as far as the ordering of $\gen^{\swdl{1}}$ and $\gen^{\swdl{2}}$ is preserved.
On the other hand, the region in between $\gen^{\swdl{1}}$ and $\gen^{\swdl{2}}$ 
can be mapped to the outside region by a suitable cyclic shift.
An apparent issue is that the inside has group-like factors 
$\gentwist^{\swdl{1/2}}$
while the outside has none.
The discrepancy is resolved by noting that the overall state should be uncharged.
This allows to apply the inverse factor $\gentwist^{\swdl{2/1}}$
to the whole state which produces precisely the desirable factors 
for the reversed bilocal contributions bounded by the insertions $\gen^{\swdl{2}}$ and $\gen^{\swdl{1}}$.
Therefore, even the bilocal contributions to an uncharged $\genyang^\twiststar$
are compatible with cyclicity.
Altogether the action of such an uncharged generator can be sketched as
\begin{align}
\numberhere
\genyang^\twiststar (\field_1\ldots\field_n) 
&=
\sum_{k=1}^n 
\field_1\ldots\field_{k-1}
(\genyang^\twiststar\field_k)
\field_{k+1}\ldots\field_n
\\
&\alignrel{}
+\sum_{j<k=1}^n 
\field_1\ldots\field_{j-1}
(\gen^{\twiststar\swdl{1}}\field_j)
\gentwist^{\swdl{1/2}}\field_{j+1}\ldots\gentwist^{\swdl{1/2}}\field_{k-1}
(\gen^{\twiststar\swdl{2}}\field_k)
\field_{k+1}\ldots\field_n,
\end{align}
where the factors in the intermediate region may be written alternatively as
$\gentwist^{\swdl{1/2}}=\gentwist^{\swdl{1}\,2}=\gentwist^{\swdl{1/2}}=\gentwist^{\swdl{2}\,-2}$.

%%%%%%%%%%%%%%%%%%%%%%%%%%%%%%%%%%%%%%%%%%%%%%%%%%%%%%%%%%%%%%%%%%%%%%%%%%%%%%%%%%%%%

\subsection{Twisted Symmetries}

We can now address the invariance statements for the twist-deformed model.
Our principal strategy is to deform the undeformed symmetry statements
with the twist operator $\twist$, for example
\[
\invact=\gen \action = 0
\eqjoin{\implies}
\invact^\twiststar 
\defeq \gen^\twiststar \action^\twiststar 
= \twist \gen \action 
= \twist \invact 
= 0.
\]
This and similar statements naturally lift from the undeformed model to the twist-deformed one
because the deformed generators $\gen^\twiststar$ are defined as
the conjugation of the undeformed generators $\gen$ with the twist operator $\twist$.
It suggests that the symmetries of the undeformed model
lift to the twist-deformed model without further ado.
However, there is a caveat: 
whereas the original invariance statement is properly cyclic,
the resulting twisted statement does not necessarily respect cyclicity.
Let us therefore elaborate further on cyclicity.

%%%%%%%%%%%%%%%%%%%%%%%%%%%%%%%%%%%%%%%%
\paragraph{Twisted Cyclicity.}

Many of the invariance relations discussed in \secref{sec:forms}
rely on the cyclic shift operator $\shift$;
in particular, it arises in situations
where a generator changes the number of fields.
However, as discussed above, the twist operator $\twist$ does not respect cyclicity in general.
More concretely, $\shift$ commutes with $\twist$ if and only if 
it acts on the subspace of uncharged states.

In order to deal with cyclicity in the twist-deformed model,
we introduce the \emph{twisted cyclic shift operator}
\[
\shift_\twiststar \defeq \twist \shift \twist^{-1} . 
\]
Its action on polynomials is a cyclic shift with an additional twist factor:
\[
\label{eq:TwistedCyclicShift}
\shift_\twiststar(\field_1\ldots \field_n) 
= \exp\brk*{2 \iunit\gamma_{ab}\chargecart^a[\field_2\ldots \field_n]\chargecart^b[\field_1]} (\field_2\ldots \field_n \field_1).
\]
On polynomials which are uncharged with respect to the twist, 
the twisted cyclic shift operator reduces to the plain cyclic shift operator
because $\gamma_{ab}\chargecart^a[\field_2\ldots \field_n]=-\gamma_{ab}\chargecart^a[\field_1]$ and 
because $\gamma_{ab}$ is an anti-symmetric matrix.

Furthermore, we introduce the \emph{twisted trace} $\tr_\twiststar$
\[
\tr_\twiststar\defeq \twist\tr\twist^{-1}.
\]
The twisted trace projects to polynomials
which are invariant under the twisted cyclic shift operator. 
It can be written as the twisted cyclic average
\[
\label{eq:DefTwistTrace}
\tr_\twiststar\oper{O}_{\nfield{n}} = \frac{1}{n} \sum_{j=1}^n\shift_\twiststar^j\oper{O}_{\nfield{n}}.
\]
On uncharged polynomials the twisted trace reduces to the untwisted trace 
because the twisted cyclic shift operator reduces to the plain cyclic shift operator. 

%%%%%%%%%%%%%%%%%%%%%%%%%%%%%%%%%%%%%%%%
\paragraph{Level-Zero Generators.}

We can now discuss the invariance statements in more detail;
let us start with the level-zero generator $\gen$.

First, we consider the covariance statement of the equations of motion.
The twist deformation of the undeformed statement $\inveom_1=0$ 
with $\inveom_1$ given in \eqref{eq:Def_Y1_StrongInvariance} reads
$\inveom_1^\twiststar=0$ with 
\begin{align}
\label{eq:twistedcovariance}
\inveom_1^\twiststar\defeq\twist\inveom_1
&=
\sum_n\sum_{m=0}^{n-2}\sum_{j=2}^{n-m}
\gentwist^{-1}_{1,j-1}\.
\gen^\twiststar_{\nfield{m},j}\.
\gentwist_{j+1,n-m}\.
\action^\twiststar_{\nfield{n-m}}
\\
&\alignrel{}
+
\sum_n\sum_{m=0}^{n-2}\sum_{j=1}^{m+1}
\shift^{j-1}_\twiststar\gen^\twiststar_{\nfield{m},1}\.
\gentwist_{1}^{-1}\.
\action^\twiststar_{\nfield{n-m}}
.
\end{align}
Note that the relation takes the form of a covariance statement
for the equations of motion as in \eqref{eq:eomsym}
both for uncharged as well as for charged generators $\gen$:
For an uncharged generator with $\gentwist=\idop$,
it reduces to a local form which is equivalent to the undeformed statement \eqref{eq:Def_Y1_StrongInvariance}.
For a charged generator, the non-local factors $\gentwist$ remain.
Nevertheless, the former term still serves as the symmetry variation 
of the equations of motion, $\gen^\twiststar \eomfor{\field}^\twiststar$.
In the latter term all operators (including the twisted shift)
still act merely on the first field of the action leaving all other fields untouched.
Consequently, the remaining fields represent some combination
of the equation of motion terms $\eomfor{\field}'^{\twiststar}$.
We thus observe that covariance of the equations of motion holds
for all generators $\gen$ irrespectively of whether it is charged or not.
However, the form of the statements differs somewhat;
for uncharged generators they are perfectly local while for charged
generators the twist-deformation introduces some non-locality.

For an invariance statement of the action, we should convert $\inveom_1^{\twiststar}$
to a one-form $\inveom^{\twiststar}$ and integrate it to a zero-form $\invact^{\twiststar}$.
Here, it turns out that it makes a difference whether the underlying generator $\gen$
is charged or uncharged. For an uncharged $\gen$, we directly obtain
\[
\inveom^{\twiststar}=\tr \var_1 \inveom^{\twiststar}_1=\twist\inveom=\var\invact^{\twiststar},
\eqsep
\invact^{\twiststar}=\twist\invact.
\]
A key insight is that the trace, twist and variation operators all commute on uncharged states,
and therefore the resulting term $\inveom^{\twiststar}$ is both cyclic and closed.
It can thus be integrated to a zero-form $\invact^{\twiststar}$
which is again cyclic.

The situation is fundamentally different if $\gen$ is charged
because the trace and twist operators do not commute anymore.
In this case, we may construct a manifestly closed term
\[
\inveom^{\twiststar}_{\textnormal{closed}}=\tr_\twiststar \var_1 \inveom^{\twiststar}_1=\twist\inveom=\var\invact^{\twiststar},
\eqsep
\invact^{\twiststar}=\twist\invact.
\]
Clearly, the covariance statements $\inveom^{\twiststar}_1=0$ 
and $\inveom^{\twiststar}_{\textnormal{closed}}=0$ are equivalent.
Both terms $\inveom^{\twiststar}_{\textnormal{closed}}$ and $\invact^{\twiststar}$
are obtained by acting with the twist operator $\twist$
on the corresponding undeformed term. 
However, as $\inveom^{\twiststar}_1$ is charged, the closed form cannot be plain cyclic, 
it is rather twisted cyclic, and the same applies to its integral $\invact^{\twiststar}$.
Unfortunately, a covariance statement based on a non-cyclic polynomial
is of limited use for physics 
because it cannot be formulated as a proper field theoretic local operator.
A gauge-invariant local operator must be a trace of a product of matrix-valued fields
and will thus be manifestly cyclic whereas $\invact^{\twiststar}$ is not plain but twisted cyclic.
The term $\invact^{\twiststar}$ therefore cannot serve as 
the divergence of a Noether current. Neither can it be used towards
formulating invariance statements within quantum correlators insertion of $\invact^{\twiststar}$.
We conclude that, even though the statement $\invact^{\twiststar}=0$ holds,
this does not imply a proper symmetry of the physical model.
Nevertheless, we emphasise that the covariance statement for the equations of motion
may conceivably have relevant consequences in classical and/or quantum field theory
even for charged generators $\gen$, but these would be beyond the ordinary symmetry relations.

To circumvent the above issue, we may instead construct a manifestly plain cyclic term
\[
\inveom^{\twiststar}_{\textnormal{cyclic}}=\tr \var_1 \inveom^{\twiststar}_1.
\]
Again, the statements $\inveom^{\twiststar}_{\textnormal{cyclic}}=0$
and $\inveom^{\twiststar}_1=0$ are equivalent.
However, the plain cyclic term turns out not to be closed
as can be established by explicit calculation
\begin{align}
\var\inveom^{\twiststar}_{\textnormal{cyclic}}
&= 
-\tr \sum_{j=2}^{n-m}\sum_{k=2}^{j-1}
\brk*{1-\gentwist^2_{1,{k-1}}}
\gentwist^{-1}_{1,j-1} 
\gentwist_{j+1,n-m} \var_1\var_k\gen^{\twiststar}_{\nfield{m}j}  \action^{\twiststar}_{\nfield{n-m}}
\\
&\alignrel{}
-\tr \sum_{j=2}^{n-m}\sum_{k=j}^{j+m} 
\brk*{1-\gentwist_{1,j-1}^2\gentwist_{j,k-1}}
\var_1\var_k\gen^{\twiststar}_{\nfield{m},j} \gentwist_{1,j-1}^{-1} \gentwist_{j+1,n-m} \action^{\twiststar}_{\nfield{n-m}}
\\
&\alignrel{}
-\tr
\sum_{j=1}^{m}\sum_{k=j+1}^{m+1} \brk*{ \gentwist_{1,j-1}^2-\gentwist_{2,k-1}^2 }\var_j\var_k\gen^{\twiststar}_{\nfield{m},1}\gentwist^{-1}_1 \action^{\twiststar}_{\nfield{n-m}}
.
\end{align}
The explicit evaluation of this term is provided in \appref{sec:explicit}.

The terms in the sums are linearly independent and therefore, 
the expression only vanishes if all terms in the sums vanish. 
For general charged generators, this is not the case, 
however, for uncharged generators 
the deformation generator $\gentwist=\idop$ is the identity map 
and $\inveom_{\textnormal{cyclic}}^\twiststar$ is indeed closed.

Consequently, the term cannot be integrated properly,
and there is no suitable term $\invact^{\twiststar}$
for a plain cyclic invariance of the action statement.

%%%%%%%%%%%%%%%%%%%%%%%%%%%%%%%%%%%%%%%%
\paragraph{Level-One Generators.}

The conclusions we have drawn for level-zero generators $\gen$ 
equally apply to the local contributions to level-one generators $\genyang$.
We can thus focus our attention on the bilocal contributions to $\genyang$ 
which may or may not change some conclusions. 

As for uncharged level-zero generators $\gen$,
the uncharged level-one counterparts $\genyang$ have
proper covariance statements for the twist-deformed equations of motion as well as 
proper invariance statements for the twist-deformed action
\[
\inveomyang_1^\twiststar=0,
\eqsep
\invactyang^\twiststar=0.
\]
These are obtained by twisting the corresponding polynomials for the undeformed model
\[
\inveomyang_1^\twiststar=\twist \inveomyang_1,
\eqsep
\invactyang^\twiststar=\twist\invactyang,
\]
and the deformed statements follow from their undeformed counterparts without further ado.
The twisted invariance term $\invactyang^\twiststar=\twist\invactyang$ takes the form
\begin{align}
\label{eq:Level1SymmetryTwistedAction}
\numberhere
\invactyang^\twiststar
&=
\sum_n\sum_{m=0}^{n-2}
\tr\genyang^\twiststar_{\nfield{m},1}
\action^\twiststar_{\nfield{n-m}}
\\
&\alignrel{}
+\sum_n\sum_{m=0}^{n-2}
\sum_{l=0}^{m}
\sum_{k=2}^{n-m} \frac{k-\half n+\half m-1}{n}
\tr\gen^{\twiststar\swdl{1}}_{\nfield{m-l},k+l}\.
\gen^{\twiststar\swdl{2}}_{\nfield{l},1}\.
\gentwist^{\swdl{2}}_{2,k-1}
\gentwist^{\swdl{1}}_{k+1,n-m}\.
\action^\twiststar_{\nfield{n-m}}
\\
&\alignrel{}
+\sum_n\sum_{m=0}^{n-2}
\sum_{l=0}^{m}
\sum_{k=1}^{l+1} \frac{2k-l-2}{n}
\tr\gen^{\twiststar\swdl{1}}_{\nfield{m-l},k}\.
\gentwist^{\swdl{2}}_{1,k-1}
\gentwist^{\swdl{1}}_{k+1,l+1}\.
\gen^{\twiststar\swdl{2}}_{\nfield{l},1}
\action^\twiststar_{\nfield{n-m}}.
\end{align}

These considerations also extend to charged level-one generators,
however, as before, an invariance statement for the action, $\invactyang^\twiststar=0$,
will be of limited use in physics because 
the polynomial $\invactyang^\twiststar$ is twisted cyclic rather than cyclic.
Therefore, a charged $\genyang$ can hardly be called a symmetry of the deformed model.
Towards covariance of the equations of motion,
we should demonstrate that
\begin{align}
\numberhere
\inveomyang_1^\twiststar
&=
\sum_n
\sum_{m=0}^{n-2}
\sum_{j=2}^{n-m}
\genyang_{\nfield{m},j}^\twiststar
\gentwistyang^{-1}_{1,j-1}
\gentwistyang_{j+1,n}
\action_{\nfield{n-m}}^\twiststar
+
\sum_n
\sum_{m=0}^{n-2}
\sum_{j=1}^{m+1}
\shift_\twiststar^{j-1}
\genyang_{\nfield{m},1}^\twiststar
\gentwistyang_1^{-1}
\action_{\nfield{n-m}}^\twiststar
\\
&\alignrel{}
+
\sum_n
\sum_{m=0}^{n-2}
\sum_{l=0}^{m}
\sum_{j=2}^{n-m-1}\sum_{k=j+1}^{n-m}
\gen^{\twiststar\swdl{1}}_{\nfield{l},j}
\gen^{\twiststar\swdl{2}}_{\nfield{m-l},k}
\gentwistyang^{-1}_{1,j-1}
\gentwist^{\swdl{2}\,-1}_{j}
\gentwist^{\swdl{1/2}}_{j+1,k-1}
\gentwist^{\swdl{1}}_{k}
\gentwistyang_{k+1,n-m}
\action_{\nfield{n-m}}^\twiststar
\\
&\alignrel{}
+
\sum_n
\sum_{m=0}^{n-2}
\sum_{l=0}^{m}
\sum_{k=1}^{l+1}
\brk[s]*{
\sum_{j=1}^{k-1}
-
\sum_{j=k+m-l+1}^{m+1}
}
%\\
%&\alignrel{}\hspace{4cm}
\shift_\twiststar^{j-1}
\gen^{\twiststar\swdl{1}}_{\nfield{m-l},k}
\gentwist^{\swdl{1}\,-1}_{1,k-1}
\gentwist^{\swdl{1}}_{k+1,l+1}
\gen^{\twiststar\swdl{2}}_{\nfield{l},1}
\gentwistyang^{-1}_1
\action_{\nfield{n-m}}^\twiststar
\end{align}
consists of the twisted level-one variation of the twisted equations of motion
$\genyang^\twiststar\eomfor{\field}^\twiststar$ and a combination of 
equation of motion terms $\eomfor{\field}'^{\twiststar}$.
Showing this is directly equivalent to the derivation at level zero
in \eqref{eq:twistedcovariance}:
First, the symmetry variation terms are straight-forwardly twisted by the twist operator $\twist$
and they yield $\genyang^\twiststar\eomfor{\field}^\twiststar$. 
Second, the terms that are proportional to the equations of motion 
are of a form where there is only an operator acting on site 1 of the action.
Applying the twist to these terms can be phrased as a twist of the operator
applied to site 1 of the twisted action. This again represents 
a term which is proportional to the twisted equations of motion.

Let us add some further remarks:
First, we note that the invariance term $\invactyang^\twiststar$ in \eqref{eq:Level1SymmetryTwistedAction}
has bilocal insertions of level-zero generators $\gen^{\swdl{1}}$ and $\gen^{\swdl{2}}$
which are supplemented by non-local insertions of $\gentwist^{\swdl{2}}=\gentwist^{\swdl{1}\,-1}$.
These non-local deformations are typical of twist deformations,
and they apply only to contributions with charged level-zero generators $\gen$.

This points at an important feature of Yangian symmetry for twist deformations:
Above we have seen that we should consider only such level-zero and level-one generators
which are uncharged under the twist deformation.
However, in order to formulate the bilocal terms in the level-one action of $\genyang$,
we still need all uncharged and charged level-zero generators $\gen$.
The residual symmetry algebra is therefore not quite the Yangian extension
of the uncharged algebra at level-zero,
but the full twist-deformed Yangian is required
to properly express the relevant symmetries.
In fact, the bilocal terms with uncharged level-zero generators $\gen^{\swdl{1}}$ and $\gen^{\swdl{2}}$
have no further non-local deformations; these arise only if the level-zero generators are charged,
and consequently do not belong to the Yangian extension of the level-zero subalgebra.

We also point out a related observation regarding the relationship 
between covariance statements for the equations of motion and 
invariance statements for the action at level one:
In the undeformed case, this relationship was
$\var\invactyang=\inveomyang_{\textnormal{eom}}+\inveomyang_{\mathrm{\Delta}}$
with $\inveomyang_{\textnormal{eom}}=\tr \var_1\inveomyang_1$
and $\inveomyang_{\mathrm{\Delta}}$ a term which must be identically zero
in any model with level-zero invariance.
For uncharged level-one generators $\genyang$,
we find the corresponding relationship
\[
\var\invactyang^{\twiststar}=\inveomyang^{\twiststar}_{\textnormal{eom}}+\inveomyang_{\mathrm{\Delta}}^{\twiststar},
\eqsep
\inveomyang_{\mathrm{\Delta}}^{\twiststar}=\twist \inveomyang_{\mathrm{\Delta}}.
\]
Here, the difference term takes the form
\begin{align}
\numberhere
\inveomyang_{\mathrm\Delta}^\twiststar
&=
-\sum_n\sum_{m=0}^{n-2}
\tr_\twiststar 
\brk!{\gen^{\twiststar\swdl{1}}_{\nfield{m},1}
\gentwist^{\swdl{1}\,-1}_1
\var_1\invact^{\twiststar\swdl{2}}_{\nfield{n-m}}
  -\gen[H]^\twiststar_{\nfield{m},1}\var_1\action_{\nfield{n-m}}^\twiststar}
\\
&\alignrel
+\sum_n\sum_{m=0}^{n-2}
\sum_{j=1}^{n}
\frac{m+2-2j}{n}
\tr_\twiststar \var_{j}
\brk!{\gen^{\twiststar\swdl{1}}_{\nfield{m},1}
\gentwist^{\swdl{1}\,-1}_1
\invact^{\swdl{2}\twiststar}_{\nfield{n-m}}
  -\gen[H]^\twiststar_{\nfield{m},1}\action_{\nfield{n-m}}^\twiststar} .
\end{align}
We note that $\inveomyang_{\mathrm{\Delta}}^{\twiststar}$ is only zero
if there are invariances for all level-zero generators $\gen$
including the charged ones.
Here, our above conclusion that only the uncharged generators actually provide symmetries of the action
appears contradictory.
Nevertheless, it is consistent:
We actually only need covariance of the equations of motion to establish
that $\inveomyang_{\mathrm{\Delta}}^{\twiststar}$ is zero,
and this also holds for the charged level-zero generators.
This insight can be related to 
the second insertion of $\gen$ which implies a variation that turns the action into the equations of motion.
The reference to covariance of the equations of motion also makes sense in view 
of the fact that we are discussing a relation which involves the equations of motion in the first place.

%%%%%%%%%%%%%%%%%%%%%%%%%%%%%%%%%%%%%%%%
\paragraph{Summary.}

Altogether, we have carried over the level-zero and level-one invariance statements
from the undeformed model to the deformed model by transforming them with 
the twist operator $\twist$.
This works well for all generators on a technical level,
however, we observe a relevant difference
between covariance of the equations of motion and invariance of the action:
The former has no cyclic structure and covariance holds for all generators $\gen$ and $\genyang$ alike.
The situation is more restricted for invariance of the action 
because twisting of the undeformed statement naturally leads to a twisted cyclic polynomial,
whereas a proper field theory invariance statement would require a plain cyclic polynomial instead.
In the case of uncharged generators $\gen$ and $\genyang$,
the twisted cyclicity of the polynomial reduces to plain cyclicity 
thus yielding a proper invariance statement for field theory.
Conversely, for generators $\gen$ and $\genyang$ which are charged under the twist,
the resulting terms are not plain cyclic and do not lead to proper symmetries.
An alternative approach to compose the covariance statement of the equations of motion 
in a manifestly cyclic way suffers from the resulting one-form not being closed.
Likewise it cannot be integrated to a proper invariance statement for the action.

%%%%%%%%%%%%%%%%%%%%%%%%%%%%%%%%%%%%%%%%
\paragraph{Resulting Symmetries.}

If we adhere to the notion that the symmetry algebra consists of invariances of the action,
only the subalgebra of generators which are uncharged under the twist provides symmetries.
In the case of gamma-deformed $\superN=4$ SYM, 
the $\alg{psu}(2,2|4)$ subalgebra of uncharged generators is $\alg{su}(2,2)\times\alg{u}(1)^3$;
for beta-deformed $\superN=4$ SYM, it extends to $\alg{su}(2,2|1)\times\alg{u}(1)^2$.
Together with the corresponding level-one generators we obtain a quantum algebra of Yangian type.
We emphasise that the resulting quantum algebra is not the plain Yangian 
of the level-zero subalgebra, but elements of the underlying $\alg{psu}(2,2|4)$ Yangian remain:
In particular, the bilocal action of level-one generators $\genyang$
not only relies on level-zero generators from the subalgebra,
but rather requires all generators of $\alg{psu}(2,2|4)$ including the supercharges.
At the level of mathematics,
the coproduct involves generators from the full $\alg{psu}(2,2|4)$ algebra.
Therefore, the restriction of the $\alg{psu}(2,2|4)$ Yangian 
to the Yangian of the subalgebra does not close at the level of the coalgebra.
\unskip\footnote{We note that the dual Coxeter number
for the subalgebra is not zero
and would cause further conflicts with cyclicity.
The additional generators from the full $\alg{psu}(2,2|4)$ algebra
are essential for the proper working of the Yangian symmetry.}

%%%%%%%%%%%%%%%%%%%%%%%%%%%%%%%%%%%%%%%%%%%%%%%%%%%%%%%%%%%%%%%%%%%%%%%%%%%%%%%%
\section{Conclusions and Outlook}
\label{sec:conclusions}

In this article, we addressed the formulation of field theoretical statements of Yangian symmetry 
for planar $\superN=4$ super Yang--Mills theory 
and its beta/gamma-deformation
as a means of formalising the feature of integrability in these models.
We mainly addressed the following two aspects:

First, we put the statement of Yangian symmetry for gauge theories
in the planar limit developed in \cite{Beisert:2018zxs} on a more solid foundation
by translating the derivation to the framework of variational forms. 
In particular, we showed that the covariance statement for the equations of motion
can be expressed as a variational one-form in the space of fields.
This one-form is closed and can thus be integrated properly
to an invariance statement for the action
confirming the earlier proposal of \cite{Beisert:2018zxs}.
We emphasise that, as usual in this context,
the vanishing of the dual Coxeter number of the level-zero symmetry algebra $\alg{psu}(2,2|4)$
for $\superN=4$ super Yang--Mills theory is essential
to maintain compatibility with the cyclic nature of the trace over colour space in the action.

Second, we lifted these statements to the beta/gamma-deformation of the model
and discussed which parts of the Yangian algebra remain as symmetries.
We found that, in agreement with the proposal in \cite{Garus:2017bgl},
the deformed equations of motion remain covariant 
under the corresponding deformation of the complete Yangian algebra.
This conclusion can be reached easily by deforming the relevant symmetry statements 
by the twist operator $\twist$.
Conversely, the deformed action is not properly invariant under 
the complete Yangian algebra, but only under its uncharged generators
with respect to the twist. For such uncharged generators, 
we remain within the space of cyclic objects where field theoretical statements
on symmetry make sense.
For charged generators, however, we cannot construct a proper representation 
on cyclic objects such as the action.
Here, the discrepancy between the applicable symmetries for the action 
and for the equations of motion can be explained as follows:
The covariance statement for the equations of motion
cannot both be properly cyclic and a closed variational one-form
at the same time. Either we cannot integrate it to an invariance for the action
or it is not cyclic.
Supposing that the invariances of the action 
describe proper symmetries of the model,
we conclude that beta/gamma-deformed $\superN=4$ SYM
is symmetric under the uncharged generators of the $\alg{psu}(2,2|4)$ Yangian algebra.
We emphasise that this algebra is not exactly the Yangian extension of
the uncharged subalgebra $\alg{psu}(2,2)\times\alg{u}(1)^3$ or $\alg{psu}(2,2|1)\times\alg{u}(1)^2$ of $\alg{psu}(2,2|4)$
because even the charged generators are required to formulate the action of the level-one (and higher) Yangian generators.
For instance, bilocal combinations of the supercharges 
do arise in the level-one action of level-one $\alg{psu}(2,2)$ generators
and moreover their action is dressed by non-local contributions of group-like factors between these two.

\smallskip

We have come across some minor issues which are left to be understood better.
For one, the implications of the reduction of the $\alg{psu}(2,2|4)$ Yangian algebra
to the subset of uncharged generators is not yet very clear. 
On the one hand, it yields a proper subalgebra.
On the other hand, the coalgebra does not close,
and the charged generators still play a role in invariance statements
(even though they do not lead to invariance statements on their own).

Another remaining puzzle is presented by the overlapping terms
within the level-one covariance and invariance statements,
see the last lines of both \eqref{eq:Def_Y1h_StrongInvariance} and \eqref{eq:Xhat}.
Such terms are intrinsically based on non-linear contributions to the level-zero generators
which change the number of fields.
Thus, they are not captured and not accounted for
by the standard coalgebra structures of Yangian algebras
which only lead to local and bilocal contributions.
On the one hand, one may view overlapping terms
as an extension of the bilocal terms to accommodate for cyclicity.
On the other hand, an ab initio derivation of these terms would be desirable.
One ansatz in the new framework of variational forms
would be to consider $\genyang\var\action$ as opposed to $\inveomyang=\var\genyang\action$.
Another ansatz might be to consider a bilocal operator of the kind
“$\var\wedge\gen$”
or a trilocal operator of the kind
“$\var\wedge\genyang$” alias “$\var\wedge\gen^{\swdl{1}}\wedge\gen^{\swdl{2}}$”
which combine the Yangian algebra with the algebra of exterior derivatives.
Similar multi-local combinations of operators
were discussed in the context of gauge fixing \cite{Beisert:2018ijg}
where the role of the variation operator $\var$
was taken by the BRST symmetry generator $\gen[Q]$
both of which square to zero.

\medskip

With Yangian symmetry for gamma-deformed $\superN=4$ Yang--Mills theory under good control,
the next step is to consider the limit of the fishnet model \cite{Gurdogan:2015csr,Chicherin:2017cns}.
This is an interesting case because the fishnet model has a drastically reduced
complexity yet some amount of Yangian symmetry still applies with minor adjustments
to the representation. However, it is not clear how these adjustments
are to be applied for a proper invariance of the action statement.
The fishnet limit of the present investigations
will shed some light in terms of a constructive deduction
for the invariance statement. Here, one will have to find a way
to deal with the reduction of the field content.
The letter may well relate to and explain
the required adjustments for the Yangian representation.

%%%%%%%%%%%%%%%%%%%%%%%%%%%%%%%%%%%%%%%%%%%%%%%%%%%%%%%%%%%%%%%%%%%%%%%%%%%%%%%%
\pdfbookmark[1]{Acknowledgements}{ack}
\section*{Acknowledgements}

We thank Axel Kleinschmidt for discussions related to the work.
The work of NB is partially supported 
by the Swiss National Science Foundation
through the NCCR SwissMAP. 
BK thanks the Institute for Theoretical Physics at ETH Zürich for the repeated hospitality. 
The work of BK is partially supported by the Konrad-Adenauer-Stiftung.

%%%%%%%%%%%%%%%%%%%%%%%%%%%%%%%%%%%%%%%%%%%%%%%%%%%%%%%%%%%%%%%%%%%%%%%%%%%%%%%%
%%%%%%%%%%%%%%%%%%%%%%%%%%%%%%%%%%%%%%%%%%%%%%%%%%%%%%%%%%%%%%%%%%%%%%%%%%%%%%%%
\appendix

%%%%%%%%%%%%%%%%%%%%%%%%%%%%%%%%%%%%%%%%%%%%%%%%%%%%%%%%%%%%%%%%%%%%%%%%%%%%%%%%
\section{Explicit calculations}
\label{sec:explicit}

The variation of
\[
\inveom^{\twiststar}_{\textnormal{cyclic}}
=
\tr \var_1 \inveom^{\twiststar}_1
=
\tr \sum_{j=2}^{n-m} \var_1\gen^{\twiststar}_{\nfield{m}j} \gentwist_{1,j-1}^{-1} \gentwist_{j+1,n-m} \action^{\twiststar}_{\nfield{n-m}}
+
\tr\sum_{j=1}^{m+1} \var_1 \shift_{\twiststar}^{j-1} \gen^{\twiststar}_{\nfield{m}1} \gentwist_1^{-1} \action^{\twiststar}_{\nfield{n-m}}
\]
is 
\begin{align}
\label{eq:VarGenTwistCyclic}
\var \inveom^{\twiststar}_{\textnormal{cyclic}}
&=
-\tr \Bigg[
\sum_{j=2}^{n-m}\sum_{k=2}^n \var_1\var_k\gen^{\twiststar}_{\nfield{m}j} \gentwist_{1,j-1}^{-1} \gentwist_{j+1,n-m} \action^{\twiststar}_{\nfield{n-m}}
\\
&\qquad\qquad
+\sum_{j=1}^{m+1}\sum_{k=2}^n \var_1\var_k \shift_{\twiststar}^{j-1} \gen^{\twiststar}_{\nfield{m}1} \gentwist^{-1}_1 \action^{\twiststar}_{\nfield{n-m}}
\Bigg],
\end{align}
where we used $\var^2 = 0$. The expression $\var\inveom^{\twiststar}_{\textnormal{cyclic}}$ has three different contributions, which are in general linearly independent. We call $I_i$ with $i = 0,1,2$ the contribution, where exactly $i$ number of variations $\var$ act on the image of $\gen^{\twiststar}_{\nfield{m}}$.

We evaluate first $I_0$, where the variations $\var$ do not act on the image of $\gen^{\twiststar}$, which has therefore only contributions from the first term in \eqref{eq:VarGenTwistCyclic}
\begin{align}
I_0
&=
-\tr \sum_{j=2}^{n-m}\sum_{k=2}^{j-1} \var_1\var_k\gen^{\twiststar}_{\nfield{m}j} \gentwist^{-1}_{1,j-1} \gentwist_{j+1,n-m} \action^{\twiststar}_{\nfield{n-m}}
\\
&\alignrel{}
-\tr \sum_{j=2}^{n-m}\sum_{k=j+m+1}^n \var_1\var_k\gen^{\twiststar}_{\nfield{m}j} \gentwist^{-1}_{1,j-1} \gentwist_{j+1,n-m} \action^{\twiststar}_{\nfield{n-m}}.
\end{align}
In the second line, the variation $\var_k$ acts to the right of $\gen^{\twiststar}$. 
To compare this expressions to the first line, we use cyclicity of the trace and cyclic shifts, which of course also shifts the twist generator $\gentwist$, 
such that it acts on the same fields but on the shifted position. 
After readjusting the limits of the sum the two terms become
\[
I_0
=
-\tr \sum_{j=2}^{n-m}\sum_{k=2}^{j-1}
\brk*{1-\gentwist_{1,k-1}^2}
\gentwist^{-1}_{1,j-1} 
\gentwist_{j+1,n-m} \var_1\var_k\gen^{\twiststar}_{\nfield{m}j}  \action^{\twiststar}_{\nfield{n-m}},
\]
where the second term in the bracket is the second term in the equation before. 
This can be easily verified by comparing the deformation factors acting between the variations. 
The extra squared deformation factor arises from the cyclic shift, which maps the fields to the very right with deformation factor $\gentwist$ to the very left, where the fields in the first term have the deformation factor $\gentwist^{-1}$.

The term $I_1$ has contributions from the first and the second term in \eqref{eq:VarGenTwistCyclic}
\begin{align}
I_1
&=
-\tr \sum_{j=2}^{n-m}\sum_{k=j}^{j+m} \var_1\var_k\gen^{\twiststar}_{\nfield{m}j} \gentwist^{-1}_{1,j-1} \gentwist_{j+1,n-m} \action^{\twiststar}_{\nfield{n-m}}
\\
&\alignrel{} -\tr \sum_{j=1}^{m+1}\sum_{k=1}^{m+1} \var_1\shift_{\twiststar}^{j-1}\var_k\gen^{\twiststar}_{\nfield{m},1} \gentwist_{2,n-m} \action^{\twiststar}_{\nfield{n-m}}.
\end{align}
Again, we use cyclicity of the trace $\tr$ to bring the field polynomials in the second term in the same cyclic order as the first term
\[
I_1
=
-\tr \sum_{j=2}^{n-m}\sum_{k=j}^{j+m} 
\brk*{1-\gentwist_{1,j-1}^2\gentwist_{j,k-1}}
\var_1\var_k\gen^{\twiststar}_{\nfield{m}j} \gentwist^{-1}_{1,j-1} \gentwist_{j+1,n-m} \action^{\twiststar}_{\nfield{n-m}}.
\]
The second term in the bracket with the deformation generator $\gentwist_{1,j-1}^2\gentwist_{j,k-1}$ is the second line in the equation before. 
The factor $\gentwist_{1,j-1}^2$ arises from the cyclic shift analogous to the case of $I_0$ and the second factor $\gentwist_{j,k-1}$ comes from the cyclic shift $\shift_{\twiststar}$ above.

Only the second term in \eqref{eq:VarGenTwistCyclic} contributes to $I_2$ and we obtain
\begin{align}
I_2
&=
-\tr
\sum_{j=1}^{m+1}\sum_{k=1}^{m+1} \var_1\shift_{\twiststar}^{j-1}\var_k\gen^{\twiststar}_{\nfield{m}1}\gentwist^{-1}_1 \action^{\twiststar}_{\nfield{n-m}}
\\
%\aeq
%-\tr
%\sum_{j=1}^{m}\sum_{k=j+1}^{m+1} \shift_{\twiststar}^{j-1}\var_j\var_k\gen^{\twiststar}_{\nfield{m}1}\gentwist^{-1}_1 \action^{\twiststar}_{\nfield{n-m}}
%+
%\tr
%\sum_{j=2}^{m+1}\sum_{k=1}^{j-1} \shift_{\twiststar}^{j-1}\var_j\var_k\gen^{\twiststar}_{\nfield{m}1}\gentwist^{-1}_1 \action^{\twiststar}_{\nfield{n-m}}
%\\
&=
-\tr
\sum_{j=1}^{m}\sum_{k=j+1}^{m+1} \shift_{\twiststar}^{j-1}\var_j\var_k\gen^{\twiststar}_{\nfield{m}1}\gentwist^{-1}_1 \action^{\twiststar}_{\nfield{n-m}}
+
\tr
\sum_{j=1}^{m}\sum_{k=j+1}^{m+1} \shift_{\twiststar}^{k-1}\var_j\var_k\gen^{\twiststar}_{\nfield{m}1}\gentwist^{-1}_1 \action^{\twiststar}_{\nfield{n-m}}
\\
&=
-\tr
\sum_{j=1}^{m}\sum_{k=j+1}^{m+1} \brk*{\gentwist_{1,j-1}^2-\gentwist_{2,k-1}^2 }\var_j\var_k\gen^{\twiststar}_{\nfield{m}1}\gentwist^{-1}_1 \action^{\twiststar}_{\nfield{n-m}}.
\end{align}
In the second line we permuted the variation and the cyclic shift and swapped the indices of the sums $j\leftrightarrow k$. In the last line we evaluated the deformation factor of the twisted cyclic shift operator and absorbed the untwisted cyclic shift operator in the cyclicity of the trace.

%%%%%%%%%%%%%%%%%%%%%%%%%%%%%%%%%%%%%%%%%%%%%%%%%%%%%%%%%%%%%%%%%%%%%%%%%%%%%%%%
%%%%%%%%%%%%%%%%%%%%%%%%%%%%%%%%%%%%%%%%%%%%%%%%%%%%%%%%%%%%%%%%%%%%%%%%%%%%%%%%
\begin{bibtex}[\jobname]

@article{Beisert:2017pnr,
    author = "Beisert, Niklas and Garus, Aleksander and Rosso, Matteo",
    title = "{Yangian Symmetry and Integrability of Planar N=4 Supersymmetric Yang-Mills Theory}",
    eprint = "1701.09162",
    archivePrefix = "arXiv",
    primaryClass = "hep-th",
    reportNumber = "HU-EP-17-03",
    doi = "10.1103/PhysRevLett.118.141603",
    journal = "Phys. Rev. Lett.",
    volume = "118",
    number = "14",
    pages = "141603",
    year = "2017"
}

@article{Beisert:2018zxs,
    author = "Beisert, Niklas and Garus, Aleksander and Rosso, Matteo",
    title = "{Yangian Symmetry for the Action of Planar N=4 Super Yang-Mills and N=6 Super Chern-Simons Theories}",
    eprint = "1803.06310",
    archivePrefix = "arXiv",
    primaryClass = "hep-th",
    reportNumber = "HU-EP-18/08, HU-EP-18-08",
    doi = "10.1103/PhysRevD.98.046006",
    journal = "Phys. Rev. D",
    volume = "98",
    number = "4",
    pages = "046006",
    year = "2018"
}

@article{Beisert:2018ijg,
    author = "Beisert, Niklas and Garus, Aleksander",
    title = "{Yangian Algebra and Correlation Functions in Planar Gauge Theories}",
    eprint = "1804.09110",
    archivePrefix = "arXiv",
    primaryClass = "hep-th",
    doi = "10.21468/SciPostPhys.5.2.018",
    journal = "SciPost Phys.",
    volume = "5",
    number = "2",
    pages = "018",
    year = "2018"
}

@article{Garus:2017bgl,
    author = "Garus, Aleksander",
    title = "{Untwisting the symmetries of $\beta$-deformed Super-Yang--Mills}",
    eprint = "1707.04128",
    archivePrefix = "arXiv",
    primaryClass = "hep-th",
    doi = "10.1007/JHEP10(2017)007",
    journal = "JHEP",
    volume = "10",
    pages = "007",
    year = "2017"
}

@article{Drummond:2009fd,
    author = "Drummond, James M. and Henn, Johannes M. and Plefka, Jan",
    editor = "Liu, Feng and Xiao, Zhigang and Zhuang, Pengfei",
    title = "{Yangian symmetry of scattering amplitudes in N=4 super Yang-Mills theory}",
    eprint = "0902.2987",
    archivePrefix = "arXiv",
    primaryClass = "hep-th",
    reportNumber = "HU-EP-09-06, LAPTH-1308-09",
    doi = "10.1088/1126-6708/2009/05/046",
    journal = "JHEP",
    volume = "05",
    pages = "046",
    year = "2009"
}

@article{ReshetikhinTwist,
     author    = "Reshetikhin, N. {\relax Yu}.",
     title     = "Multiparametric quantum groups and twisted quasitriangular
                  Hopf algebras",
     journal   = "Lett. Math. Phys.",
     volume    = "20",
     year      = "1990",
     pages     = "331",
     doi = "10.1007/BF00626530"
}

@article{Leigh:1995ep,
    author = "Leigh, Robert G. and Strassler, Matthew J.",
    title = "{Exactly marginal operators and duality in four-dimensional N=1 supersymmetric gauge theory}",
    eprint = "hep-th/9503121",
    archivePrefix = "arXiv",
    reportNumber = "RU-95-2",
    doi = "10.1016/0550-3213(95)00261-P",
    journal = "Nucl. Phys. B",
    volume = "447",
    pages = "95--136",
    year = "1995"
}

@article{Drinfeld:1989st,
    author = "Drinfeld, V. G.",
    title = "Quasi Hopf algebras",
    journal = "Alg. Anal.",
    volume = "1",
    pages = "114--148",
    year = "1989",
    url = {https://www.mathnet.ru/eng/aa/v1/i6/p114}
}

@article{Lunin:2005jy,
    author = "Lunin, Oleg and Maldacena, Juan Martin",
    title = "{Deforming field theories with U(1) $\times$ U(1) global symmetry and their gravity duals}",
    eprint = "hep-th/0502086",
    archivePrefix = "arXiv",
    doi = "10.1088/1126-6708/2005/05/033",
    journal = "JHEP",
    volume = "05",
    pages = "033",
    year = "2005"
}

@article{Frolov:2005dj,
    author = "Frolov, Sergey",
    title = "{Lax pair for strings in Lunin-Maldacena background}",
    eprint = "hep-th/0503201",
    archivePrefix = "arXiv",
    doi = "10.1088/1126-6708/2005/05/069",
    journal = "JHEP",
    volume = "05",
    pages = "069",
    year = "2005"
}

@article{Beisert:2005if,
    author = "Beisert, N. and Roiban, R.",
    title = "{Beauty and the twist: The Bethe ansatz for twisted N=4 SYM}",
    eprint = "hep-th/0505187",
    archivePrefix = "arXiv",
    reportNumber = "PUTP-2162",
    doi = "10.1088/1126-6708/2005/08/039",
    journal = "JHEP",
    volume = "08",
    pages = "039",
    year = "2005"
}

@article{Gurdogan:2015csr,
    author = {Gürdo\u{g}an, {\relax Ö}mer and Kazakov, Vladimir},
    title = "{New Integrable 4D Quantum Field Theories from Strongly Deformed Planar N=4 Supersymmetric Yang-Mills Theory}",
    eprint = "1512.06704",
    archivePrefix = "arXiv",
    primaryClass = "hep-th",
    doi = "10.1103/PhysRevLett.117.201602",
    journal = "Phys. Rev. Lett.",
    volume = "117",
    number = "20",
    pages = "201602",
    year = "2016"
}

@article{Chicherin:2017cns,
    author = {Chicherin, Dmitry and Kazakov, Vladimir and Loebbert, Florian and Müller, Dennis and Zhong, De-liang},
    title = "{Yangian Symmetry for Bi-Scalar Loop Amplitudes}",
    eprint = "1704.01967",
    archivePrefix = "arXiv",
    primaryClass = "hep-th",
    reportNumber = "HU-EP-17-09, MITP-17-022, LPTENS-17-07",
    doi = "10.1007/JHEP05(2018)003",
    journal = "JHEP",
    volume = "05",
    pages = "003",
    year = "2018"
}

@article{Beisert:2010jr,
    author = "Beisert, Niklas and others",
    title = "{Review of AdS/CFT Integrability: An Overview}",
    eprint = "1012.3982",
    archivePrefix = "arXiv",
    primaryClass = "hep-th",
    reportNumber = "AEI-2010-175, CERN-PH-TH-2010-306, HU-EP-10-87, HU-MATH-2010-22, KCL-MTH-10-10, UMTG-270, UUITP-41-10",
    doi = "10.1007/s11005-011-0529-2",
    journal = "Lett. Math. Phys.",
    volume = "99",
    pages = "3--32",
    year = "2012"
}

@article{Drinfel'd:1985,
    author = "Drinfel'd, Vladimir Gershonovich",
    title = "Hopf algebras and the quantum Yang–Baxter equation",
    journal   = "Sov. Math. Dokl.",
    volume    = "32",
    pages     = "254-258",
    year      = "1985"
}

@article{Drinfel'd:1988,
    author = "Drinfel'd, Vladimir Gershonovich",
    title = "Quantum groups",
    doi = "10.1007/BF01247086",
    journal = "J. Sov. Math.",
    volume = "41",
    issue = "2",
    pages = "898--915",
    year = "1988"
}

@article{Dolan:2004ps,
	author    = "Dolan, Louise and Nappi, Chiara R. and Witten, Edward",
	title     = "Yangian symmetry in $D=$4 superconformal Yang--Mills theory",
	booktitle = "Quantum Theory and Symmetries",
	series    = "Proceedings of the 3rd International Symposium, Cincinnati, USA, 10-14 September 2003",
	editor    = "Argyres, P. C. and others",
	publisher = "World Scientific",
	year      = "2004",
	address   = "Singapore",
    doi = "10.1142/9789812702340_0036",
    pages = "300--315",
	eprint    = "hep-th/0401243",
	archivePrefix = "arXiv",
}

@article{Dolan:2003uh,
    author = "Dolan, Louise and Nappi, Chiara R. and Witten, Edward",
    title = "{A Relation between approaches to integrability in superconformal Yang-Mills theory}",
    eprint = "hep-th/0308089",
    archivePrefix = "arXiv",
    doi = "10.1088/1126-6708/2003/10/017",
    journal = "JHEP",
    volume = "10",
    pages = "017",
    year = "2003"
}

@article{Beisert:2006fmy,
    author = "Beisert, Niklas",
    editor = "Faddeev, L. and Henneaux, M. and Kashaev, R. and Volkov, A. and Lambert, F.",
    title = "{The S-matrix of AdS / CFT and Yangian symmetry}",
    eprint = "0704.0400",
    archivePrefix = "arXiv",
    primaryClass = "nlin.SI",
    reportNumber = "AEI-2007-019",
    doi = "10.22323/1.038.0002",
    journal = "PoS",
    volume = "SOLVAY",
    pages = "002",
    year = "2006"
}

@article{Muller:2013rta,
    author = {Müller, Dennis and Münkler, Hagen and Plefka, Jan and Pollok, Jonas and Zarembo, Konstantin},
    title = "{Yangian Symmetry of smooth Wilson Loops in N=4 super Yang-Mills Theory}",
    eprint = "1309.1676",
    archivePrefix = "arXiv",
    primaryClass = "hep-th",
    reportNumber = "HU-EP-13-42, NORDITA-2013-64, UUITP-10-13",
    doi = "10.1007/JHEP11(2013)081",
    journal = "JHEP",
    volume = "11",
    pages = "081",
    year = "2013"
}

\end{bibtex}

\bibliographystyle{nb}
\bibliography{\jobname}

\end{document}